\documentclass{iopart}
\usepackage{graphicx}
\usepackage{amssymb}
\begin{document}

\def\dirac{i\partial\!\!\!/}
\def\Dirac{D\!\!\!\!/}

\def\DDirac{i\partial\!\!\!/+eA\!\!\!/}
\def\e{{\rm e}}
\def\egamma{\buildrel-\over\gamma}
\def\eE{\buildrel-\!-\over E}
\def\ex{\buildrel-\over x}
\def\eA{\buildrel-\over A}
\def\oq{\buildrel-\over q}
\def\pa{\partial}
\def\vu{\upsilon}
\def\ve{\varepsilon}
\def\sumint{\sum\!\!\!\!\!\!\!\!\int}
\def\lbar{^-\!\!\!\!\lambda}
\def\R{\Re{\rm e}}
\def\I{\Im{\rm m}}

\newcommand{\beq}{\begin{eqnarray}}
\newcommand{\eeq}{\end{eqnarray}}
\newcommand{\nn}{\nonumber}
\newcommand{\no}{\nonumber\\}

\article[Planar QED]{}
{Planar QED at finite temperature and density: Hall conductivity, Berry's phases and minimal conductivity of graphene}

\date{\today}

\author{C.G. Beneventano}\address{Departamento de F\'{\i}sica and IFLP-
Universidad Nacional de La Plata - CONICET}
\author{Paola Giacconi}\address{Istituto Nazionale di Fisica Nucleare - Sezione di Bologna}
\author{E. M. Santangelo}\address{Departamento de F\'{\i}sica and IFLP-
Universidad Nacional de La Plata - CONICET}
\author{Roberto Soldati}\address{Dipartimento di Fisica - Universit\`a
di Bologna, Istituto Nazionale di Fisica Nucleare - Sezione di
Bologna}
\begin{abstract}

We study 1-loop effects for massless Dirac fields in two spatial dimensions, coupled to homogeneous electromagnetic backgrounds,
both at zero and at finite temperature and density. In the case of a purely magnetic field, we analyze the relationship between the invariance of the theory under large gauge transformations, the appearance of Chern-Simons terms and of different Berry's phases.
In the case of a purely electric background field, we show that the effective Lagrangian
is independent of the chemical potential and of the temperature.
More interesting: we show that the minimal conductivity,
as predicted by the quantum field theory, is the right multiple of the conductivity quantum and is, thus, consistent with the value measured for graphene, with no extra factor of $\pi$ in the denominator.

\end{abstract}

\pacs{11.10.Wx, 02.30.Sa, 73.43.-f}

\section{Introduction}\label{intro}

Planar electrodynamics, i.e. electrodynamics in two spatial dimensions, has
attracted attention for years,
mainly due to its theoretical departure from what we are used to in three
spatial dimensions -- for two reviews of the subject,
with emphasis on Chern-Simons terms, and many relevant references
see \cite{dunne,dittrich}. A pedagogical introduction to planar electrodynamics can be found in \cite{pedagogical}.

In turn, more than twenty years ago, it was shown that, in the tight-binding
approach,
the effective theory at low momenta for a two-dimensional lattice of carbon
atoms (which is now known as graphene)
is nothing but a theory of massless Dirac fields in a
{\em 2+1}-dimensional Minkowski space-time, in a reducible representation of
the Clifford algebra
(each irreducible representation corresponding to one Dirac point or valley)
and with two fermion species,
representing the spin of the interacting electrons \cite{semenoff,mele}.

Stable monolayer samples of graphene were first isolated in 2004
\cite{novo0}.
The measurements, in 2005, of the Hall conductivity in those samples confirmed
that
massless planar quantum electrodynamics is adequate to describe the effective behavior of quasi particles in graphene \cite{nature1}.
The special, unique form for the conductivity of graphene,
its many potential applications, as well as the link that graphene establishes
between a relativistic field theory and a condensed matter system have
triggered, during the last years,
a huge amount of work, both experimental and theoretical, on the subject -- for a recent review, see \cite{geim}.

In \cite{jphysa07}, we studied planar spinor electrodynamics at finite temperature and particle density,
in the case of massless Dirac quantized fields in the presence of a homogeneous magnetic background.
To this aim we evaluated,
in the zeta function regularization approach \cite{dowker},
the one loop effective action of the three-dimensional Euclidean theory, at finite temperature and in the presence of a real chemical potential.
From such a result, we obtained the mean particle density, or equivalently the net charge,
as a function of the temperature and of the chemical potential. The zero
temperature limit of this charge,
followed by a Lorentz boost, allowed us to obtain the Hall current as a
function of the chemical potential.
In particular, the behavior of the Hall conductivity around zero chemical
potential was shown, in the same reference,
to depend on the selection made for the so called ''phase" of the Dirac
determinant \cite{ecz} (for a discussion of this point, see also \cite{BS1}).
The present paper is a natural continuation of our previous work on the subject. Indeed, here we will study one-loop effective actions in the presence of those particular classical configurations characterizing the relevant experimental setups.

In sections \ref{sect1} and \ref{sect2}, we present some basic conventions and
calculations needed to study
the massless theory in a purely magnetic homogeneous background, this time for a general complex-valued chemical potential.
The introduction of an imaginary part in the chemical potential will allow us
to discuss,
in sections \ref{sect3} and \ref{sect4}, the invariance of the effective
theory under ''large" gauge transformations,
their interpretation as $4\pi$ rotations, and the relationship among different
phases of the determinant,
the appearance (or lack thereof) of geometric phases
at finite temperature, as well as their interpretation as Chern-Simons terms or Berry's phases \cite{berry} in the zero temperature limit.
These points were briefly discussed by two of us in the proceeding issue \cite{leipzig}.
Here, we present a more detailed analysis of these topics.

Sections \ref{sect5}, \ref{sect6} and \ref{sect7} treat the complementary case of a purely electric constant background field,
both at zero and non-vanishing, temperature and chemical potential. In section \ref{sect5},
we use the Bogoliubov transformation method to study the vacuum persistence probability,
determined by the imaginary part of the effective action at zero temperature,
in the {\em 2+1} dimensional Minkowski space-time. We obtain the probability of vacuum decay and show that, as expected for massless particles, it is far from being small. In section \ref{sect6},
we evaluate, through zeta function regularization, the effective action for the same background at finite temperature.
We show that, in this case, the net charge is zero at all temperatures
(equivalently, the effective action is independent of the chemical potential, whether real or
complex). We also show that, in going back to Minkowski space-time, the imaginary part
of the effective action coincides,
at all temperatures, with the one obtained in the previous section in the zero temperature case.

The most impressive result in this paper is probably the one contained in section \ref{sect7}, where we evaluate, through the introduction of an adequate Lagrange multiplier, the mean current in the presence of a constant electric field. We show that such a calculation predicts a minimal conductivity which, at variance with many of the different results presented in the literature (see, for instance, \cite{mincond,ludwig94}), is compatible with the measurements presented in \cite{geim}.
Finally, section \ref{conclusions} contains our main conclusions, while the appendix presents a short
overview of the main properties of the parabolic cylinder functions, needed to follow our calculation of the probability of vacuum decay at zero temperature.

\section{Dirac field interacting with a constant magnetic background at zero temperature}\label{sect1}

In this section, we use the 2+1 Minkowski space-time metric
$g^{\,\mu\nu}={\rm diag}\,(+,-,-)\,,$
and, throughout the paper (unless otherwise stated), we adopt natural units $\hbar=c=\tilde c=1\,,$ where $\tilde c$ denotes the ``speed of light''
for the two dimensional system. Note that, for graphene, $\tilde c$ is approximately two orders of magnitude
smaller
than the speed of light in vacuum \cite{geim,peres}.
We choose the following representation
for the Dirac matrices
\beq
\gamma^0=\sigma_3,\qquad\gamma^1=i\sigma_1\qquad {\rm and}\qquad
\gamma^2=i\sigma_2\,.\label{gammam}
\eeq
They satisfy the required properties
\beq
&&\gamma^{\,\mu\,\dagger}=\gamma^0\gamma^\mu\gamma^0\,,\quad
\gamma^0=\gamma^{\,0\,\dagger}\,,\quad \gamma^{\,k} = -\,\gamma^{\,k\,\dagger}\,,\quad
k=1,2\,;\\
&&\left\{\gamma^\mu,\gamma^\nu\right\}=
\gamma^\mu\gamma^\nu+\gamma^\nu\gamma^\mu=2g^{\,\mu\nu}\,{\mathbb I}\,.
\eeq

As is well known \cite{flekkoy}, a second, nonequivalent representation of the Clifford
algebra can be chosen, for instance,
 by reversing the sign of one of the spatial Dirac matrices.
We will always consider massless fields, two fermion species or flavors, and the direct sum of both nonequivalent representations, i.e., the effective model
for graphene \cite{semenoff,mele}.

For the representation in equation (\ref{gammam}),
the 1-particle Dirac hamiltonian operator $H_D$ is determined by the stationary
solutions of the Dirac equation
\beq
(\DDirac)\Psi(x)=\Dirac\;\Psi(x)=0\,,
\label{eqdirac}
\eeq
where $-e$ is the negative electron charge.
From the previous equation, it is easily seen that
\beq
H_D=-eA_0 - \alpha^k D_{\,k}\,,
\eeq
where
$\gamma^0\gamma^k=\alpha^k=(-\,\sigma_2\,,\,\sigma_1)\,$ and
and $D_k=i\partial_k+eA_k\,,\ k=1,2\,.$

If a uniform magnetic field is present along the positive $Oz-$axis
perpendicular to the $Oxy-$plane,
after setting
\beq
\Psi_E(x^0,{\bf x})=\e^{-\,iEx^0}\,\psi_E(x^1,x^2)={\rm e}^{-\,iEt}\,\psi_E(x,y)\,,
\eeq
and choosing the Landau gauge $A^{\,\mu}=(0,0,-\,x B)\,,$
with $B>0\,,$ we get the 1-particle Dirac hamiltonian
\beq
H_D &=& i\partial_x\,\sigma_2 -
\left(i\partial_y + eBx\right)\sigma_1\no
&=& \left\lgroup\begin{array}{cc}
0 & \partial_x - i\partial_y - eB x \\
-\partial_x - i\partial_y - eB x  & 0\\
\end{array}\right\rgroup\ .
\label{hamilton}
\eeq
We make the ansatz
$$
\psi_E(x,y)=\frac{1}{\surd{\,2\pi}}\int_{-\infty}^\infty dp\
\psi_{\,E,\,p}\,(x)\ \e^{\,i\,p\,y}\,,
$$
which allows us to determine the eigenvalues of the Hamiltonian from the set of equations

\beq
\left\lgroup\begin{array}{cc}
E  & -\,d_x - p + eB x \\
d_x - p + eB x  & E \\
\end{array}\right\rgroup\,\psi_{\,E,\,p}\,(x)\ =\ 0\,.
\label{hall}
\eeq
In order to find the energy eigenstates, it is convenient to
introduce the dimensionless coordinate

\beq
\xi=\sqrt{2eB}\left(x-\frac{p}{eB}\right)\,,\qquad d_x=\sqrt{2eB}\,d_\xi\,.
\eeq
In terms of this variable, the 1-particle Dirac hamiltonian can be written as

$$
H_D = \sqrt{2eB}\,
\left\lgroup\begin{array}{cc}
0 & \delta_+
\\
-\,\delta_-
& 0
\end{array}\right\rgroup\,,
$$
with

$$
\delta_\pm\ =\ d_\xi\,\mp\,{\xi\over2}\,,
$$
where the raising and lowering operators satisfy

\beq
\left[\,\delta_{\,\pm}\,,\,\delta_{\,\mp}\,\right]\ =\ \pm\,1\,,\qquad
\left[\,\delta_{\,\mp}\,,\,\delta_{\,\mp}\,\right]\ =\ 0\,,
\eeq
\beq
\delta_{\,+}\,|\,n\,\rangle = -\,|\,n+1\,\rangle\,,\qquad
\delta_{\,-}\,|\,n\,\rangle = n\,|\,n-1\,\rangle\,,
\eeq
\beq
\langle\,z\,|\,n\,\rangle = \phi_{\,n}(z)\,,\qquad
\langle\,n'\,|\,n\,\rangle = \delta_{\,nn'}\ .
\eeq

\medskip
Now, if we require the Dirac
Hamiltonian to be self-adjoint on the domain
$L^2({\mathbb R}^2)\,,$ we are driven to
the orthonormal set of functions
\beq
\left(n!\sqrt{2\pi}\,\right)^{-1/2}\,D_n(\xi)\
\equiv\ \phi_n(\xi)\,,\quad
n=0,1,2,\ldots\,,
\eeq
where
\beq
D_n(\xi)=2^{-n/2}\,{\rm e}^{-\xi^2/4}\,H_n(\xi/\sqrt2)\,
\eeq
are parabolic cylinder functions of integer order.
Notice that the normalization constants have been chosen to satisfy
\beq
\int_{-\infty}^{\infty}d\xi\ \phi_m(\xi)\,\phi_n(\xi)=\delta_{mn}\,,\qquad
n,m=0,1,2,\ldots\,.
\eeq

Hence, we obtain
the following complete and orthonormal set of stationary improper eigenfunctions
of the first order Dirac Hamiltonian. These comprise an infinite number of modes with zero energy,
\beq
\Psi_{\,0,\,p}\,(t,x,y) &=&
\psi_{\,0,\,p}\,({\bf x})\qquad\qquad(\rm Zero\ modes)\nn\\
&=&\left(eB\over \pi\right)^{1/4}\frac{1}{\sqrt{2\pi}}
\left\lgroup\begin{array}{c}
1\\
0\\
\end{array}\right\rgroup\
\exp\left\{i p y-\frac14\,\xi^2\right\}\,,
\eeq
and an infinite number of modes with nonvanishing energies,
\beq
\Psi_{n,\,p}^{\,(\pm)}\,(t,x,y) &=&
\psi_{n,\,p}^{\,(\pm)}\,({\bf x})\,
\exp\{\pm\,i\,E_n\,t\}\nn\\
&=& \left(eB\over 2\right)^{1/4}\frac{1}{\sqrt{2\pi}}
\exp\{\pm\,i\,E_n\,t+i\,p\,y\}\nn\\
&\times& 
\left\lgroup\begin{array}{c}
\phi_n(\xi)\ \cos\theta_{\,n}^{\,\pm}\\
\phi_{n-1}(\xi)\,\sin\theta_{\,n}^{\,\pm}\\
\end{array}\right\rgroup\,,
\eeq
corresponding to eigenenergies $\mp E_{\,n}=\mp \sqrt{2neB}\,,$
where $\tan\theta_{\,n}^{\,\pm} =\mp E_{\,n}$.

The spectrum is purely discrete and
each eigenvalue exhibits the well known  continuous
Landau degeneracy {\it per} unit area,
\beq
\Delta\ =\ \frac{eB}{2\pi}\,.
\label{landauB}
\eeq

In fact, the improper Dirac eigenspinors are normalized according to
\beq
&&\int d{\bf x}\ \psi_{\,0,\,l}^{\,\dagger}({\bf x})\,
\psi_{\,0,\,k}({\bf x})=\delta(k-l)\,,\\
&&\int d{\bf x}\ {\overline \psi}_{m,\,l}^{\,(\pm)}({\bf x})\,\gamma^0\,
\psi_{n,k}^{(\pm)}({\bf x})=
\delta_{\,mn}\,\delta(l-k)\,,
\eeq
and, as a consequence, we have
\beq
&&\int_{-\infty}^\infty dk\
\psi_{\,0,\,k}^{\,\dagger}({\bf x})\,
\psi_{\,0,\,k}({\bf x}) = \Delta\,,\nn\\
&&\int_{-\infty}^\infty dk\
{\overline \psi}_{n,\,k}^{\,(\pm)}({\bf x})\,\gamma^0\,
\psi_{n,\,k}^{(\pm)}({\bf x}) = \Delta\,,\qquad\quad
\forall\, n\in{\mathbb N}\,.
\label{band}
\eeq

The corresponding eigenfunctions in the other nonequivalent representation, where $\gamma^1 \rightarrow -\gamma^1$, are obtained through $\psi_{\,0,\,k}\rightarrow\sigma_1\psi_{\,0,\,k}$, $\psi_{\,n,\,k}\rightarrow\sigma_1\psi_{\,n,\,k}$, and the corresponding energies are the same as the ones found in the present representation. Finally, to describe graphene, an overall degeneracy of 2 (two species or "flavors") must be considered to take the spin of the electrons in the original tight binding model into account.

\section{Dirac field interacting with a constant magnetic background, at finite temperature and density}\label{sect2}

In order to consider the effect of temperature, we study the same problem as in the previous section, this time in the Euclidean three dimensional space, by considering the effective one-loop action in the path integral approach. This requires the evaluation of the determinant of the corresponding Euclidean Dirac operator, through some regularization method, which we will choose to be the zeta function one \cite{dowker}. To this end, we take the Euclidean gamma matrices in one of the two nonequivalent representations as
$$
\egamma_j\ =\ -\,i\,\gamma^{\,j}\,,\quad j=1,2\,,\qquad
\egamma_3\ =\ \gamma^0\,,
$$
so that
$$
\egamma_\mu\ =\ \sigma_\mu\,,\qquad \mu=1,2,3\,,\qquad
\{\egamma_\mu\,,\,\egamma_\nu\}=2\delta_{\mu\nu}\,{\mathbb I}\,,
$$
and set
$$
ix_0\ =\ x_3\,,\qquad x_{\,\mu} =\ (x_1,x_2,x_3)\ =\ ({\bf x},\tau)\ =\ (x,y,\tau)\,,
$$
together with
$$
A_0\ =\ iA_3\,,\qquad \bar A_{\,\mu}\ =\ ({\bf A},A_3)\ =\ (0,Bx,-\,i\,\vu/e)\,,
$$
where a generalized complex chemical potential $\vu=\mu+i\gamma$
has been introduced. Note that the presence of a real chemical potential
in the Minkowski space-time just amounts to a trivial overall shift of the energy scale. At finite temperature, instead, the derivative of the effective action with respect to the real chemical potential will give us the mean value of the net charge of the theory. The imaginary part of the chemical potential will, later on, allow us to interpret our results in terms of topological phases.

To describe graphene we treat the massless Euclidean Dirac operator
\beq
\left(\,\Dirac\,\right)_{\,\rm Eucl} &=&
\sigma_1\,(-\,i\,\pa_x) + \sigma_2\left(-\,i\,\pa_y - eBx\,\right)
+\sigma_3\left(\,-\,i\,\pa_{\,\tau} +\,i\,\vu\,\right)\no
&=& \sigma_\lambda (p_\lambda - eBx\,\delta_{2\lambda} -\gamma\,\delta_{3\lambda})+i\mu\sigma_3\ .
\label{diracB}\eeq

\medskip
In order to evaluate the functional determinant, we look for the spectral resolution of (\ref{diracB}), which is not a self-adjoint operator for $\mu=\R\,\vu\not=0\,.$ We restrict its domain to regular square summable functions
in $x\,.$ Moreover, we shall adopt antiperiodic
boundary conditions on the strip $0\le\tau\le\beta\,,$ where $\beta=1/T\,,$
(note that $k_{B}$ has been put to one in natural units). As a consequence of this antiperiodicity,
the three dimensional Euclidean space manifold is the direct product
$\mathcal M_3=\mathbb R^{\,2}\times C^{\,1}(\frac{\beta}{2\pi})\,,$
where $C^{\,n}(r)$ is the circumference of radius $r$
and integer winding number equal to $n\in\mathbb Z$. Hence,
the symmetry group of $\mathcal M_3$ becomes the
direct product $U(1)\times O(2)\,\dot=\,U(1)\times U(1)\,,$ that is the square of
the unitary, non simply connected, Abelian group $U(1)$.

To satisfy the antiperiodicity, we introduce the Matsubara angular frequencies
$$
\omega_{\,\ell}
=(2\ell+1)\,\frac{\pi}{\beta}\,,\qquad\quad \ell\in\mathbb Z\,,
$$
and propose, for the eigenfunctions of (\ref{diracB}),
\beq
\psi\,(x,y,\tau)=\frac{1}{\surd{\,2\pi\beta}}\ \sumint_{\,\;\,\ell,\,p}\
\exp\{\,i\left(\,\tau\,\omega_{\,\ell} + p\,y\,\right)\}\ \psi_{\,\ell,\,p}\,(x)\,,
\eeq
where
\[
\sumint_{\,\;\,\ell,\,p}\ \equiv\ \sum_{\ell\,=\,-\,\infty}^\infty\ \int_{-\infty}^\infty dp\,.
\]

Thus, the problem to be solved becomes
\beq
\left\lgroup
\begin{array}{cc}
\omega_{\,\ell} + i\,\vu - \lambda & -\,i\,d_{\,x} + i\,(\,eBx-p\,)\\
-\,i\,d_{\,x} - i\,(\,eBx-p\,)  & -\,\omega_{\,\ell} - i\,\vu - \lambda
\end{array}\right\rgroup\
\left\lgroup
\begin{array}{c}
\varphi_{\,\ell,\,p}\\
\chi_{\,\ell,\,p}
\end{array}\right\rgroup
=0\ .
\nn
\eeq
The resulting spectrum is complex and consists of two parts. We will call the first part asymmetric, since, given one eigenvalue $\lambda$ in this part of the spectrum, $-\lambda$ doesn't belong to the spectrum. The second part, which we will call the symmetric part of the spectrum, instead, is such that, for each eigenvalue $\lambda$ belonging to this part of the spectrum, $-\lambda$ also belongs to the spectrum. ( Note that this last part behaves as expected from the eigenvalues of a square root operator).

\begin{enumerate}
\item
The asymmetric piece of the spectrum, which is
an infinite ''tower" of states associated to the lowest Landau level in the Hamiltonian, is given by
\beq
\lambda_{\,\ell}\ =\ \omega_{\,\ell}\ +\,i\,\vu\ =\
\omega_{\,\ell}\ -\ \gamma\ + i\,\mu\,,\qquad\quad
\left(\,\ell\in \mathbb Z\,\right)
\label{asymspec}\eeq
and has a corresponding set of eigenfunctions
\beq
\Psi_{\,\ell,\,p}\,(x,y,\tau)&=&\left({eB\over\pi}\right)^\frac14
\left\lgroup
\begin{array}{c}
\e^{\,-\,\frac12\,eB\,\left(\,x-p/eB\,\right)^{\,2}}\\
0\end{array}\right\rgroup\no
&\times&
\exp\{\,i\,\tau\,\omega_{\,\ell} + i\,p\,y\}\
\left(\,2\pi\beta\,\right)^{\,-\frac12}\,.
\label{asymeigen}
\eeq
Their normalization is given by
\beq
\fl\qquad
\int_0^{\,\beta} d\tau\int_{-\infty}^\infty dx\int_{-\infty}^\infty dy\
\Psi_{\,\ell,\,p}\,(x,y,\tau)\,
\Psi^{\,\ast}_{\,\jmath,\,q}\,(x,y,\tau)=
\delta_{\,\ell\,\jmath}\;\delta\left(p-q\right)\,,
\nn
\eeq
and the degeneracy {\it per} unit area is, as in the zero temperature case,
$$
\int_0^{\,\beta} d\tau\int_{-\infty}^\infty dp\
\left|\,\Psi_{\,\ell,\,p}\,(x,y,\tau)\,\right|^{\,2}=
\frac{eB}{2\pi}=\Delta\ .
$$

\item
The symmetric part of the spectrum is related to the excited Landau levels. It is given by
\beq
\lambda\ =\ \pm\,\lambda_{\,\ell,\,n}\,,\qquad\quad
\lambda_{\,\ell,\,n}\ =\ \sqrt{\lambda^{2}_{\,\ell}\ +\ 2eB\,n}\,,
\label{symspec}\eeq
with $n\in{\mathbb N}$ and $\ell\in{\mathbb Z}\,.$ To obtain the
corresponding eigenstates, it is convenient to set
\beq
z=x\,\sqrt{2eB}\ -\ p\,\sqrt{{2\over eB}}\,,\qquad
d_{\,x}=\sqrt{2eB}\ d_{\,z}\;,
\eeq
so that
\beq
d_{\,x}\pm(\,p - eBx\,)=\sqrt{2eB}\left(\,d_{\,z}\mp{z\over2}\,\right)
\equiv \sqrt{2eB}\ \delta_{\,\pm}\;,
\eeq
where the raising-lowering operators satisfy
\beq
\left[\,\delta_{\,\pm}\,,\,\delta_{\,\mp}\,\right]\ =\ \pm\,1\,,\qquad
\left[\,\delta_{\,\mp}\,,\,\delta_{\,\mp}\,\right]\ =\ 0\,,
\eeq
\beq
\delta_{\,+}\,|\,n\,\rangle = -\,|\,n+1\,\rangle\,,\qquad
\delta_{\,-}\,|\,n\,\rangle = n\,|\,n-1\,\rangle\,,
\eeq
\beq
\langle\,z\,|\,n\,\rangle = \phi_{\,n}(z)\,,\qquad
\langle\,m\,|\,n\,\rangle = \delta_{\,mn}\;,
\eeq

$$
\phi_{\,n}(z)\equiv\left(n!\sqrt{2\pi}\,\right)^{-\,\frac12}\,D_n(z)
\,,\qquad
n=1,2,\ldots\;.
$$
Hence, if we rewrite the Euclidean Dirac operator in the form

\beq
\left(\,\Dirac\,\right)_{\,\rm Eucl}=\sqrt{\,2eB}\,
\left\lgroup
\begin{array}{cc}
\lbar_{\,\ell} & -\,i\,\delta_{\,+}\\
-\,i\,\delta_{\,-}  & -\,\lbar_{\,\ell}
\end{array}\right\rgroup\;,
\eeq
where $\lbar_{\,\ell}\ =\ \lambda_{\,\ell}/\sqrt{2eB}\,,$
it turns out that the
normalized eigenstates are given by

\beq
\Psi_{\ell,\,p,\,n,\,\pm}&=& \left(\,2eB\,\right)^\frac14\
\exp\{\,i\tau\,\omega_{\,\ell} + i\,p\,y\}\
\left(\,4\pi\beta\,\right)^{\,-\frac12}\no
&\times& \left(\,1+|\,\varkappa_{\,\ell,\,n\,,\,\pm}\,|^{\,2}\,\right)^{-\,\frac12}
\left\lgroup\begin{array}{c}
\pm\,i\,\phi_{\,n}(z)\\
\varkappa_{\,\ell,\,n\,,\,\pm}\,\phi_{\,n-1}(z)\\
\end{array}\right\rgroup\;,
\eeq
with

\beq
\varkappa_{\,\ell,\,n\,,\,\pm} =
\frac{1}{\surd{\,2eB}}\,\left(\,\lambda_{\,\ell,\,n}\,\mp\,\lambda_{\,\ell}\,\right)
=\frac{n\,\sqrt{\,2eB}}{\lambda_{\,\ell,\,n}\,\pm\,\lambda_{\,\ell}}\;.
\nn
\eeq
Moreover, by setting
$$
\lambda_{\,\ell,\,n}\equiv\rho_{\,\ell,\,n}\left(\,\cos\alpha_{\,\ell,\,n}
+ i\,\sin\alpha_{\,\ell,\,n}\,\right)\;,
$$
one finds
$$
\rho_{\,\ell,\,n}=\frac{1}{\surd{\,2eB}}\left\lbrace
\left[\,\left(\,\omega_{\,\ell}-\gamma\,\right)^2-\mu^2+2eB\,n\,\right]^2+
4\mu^2\left(\,\omega_{\,\ell}-\gamma\,\right)^2\right\rbrace^\frac14\;,
$$
$$
\tan2\alpha_{\,\ell,\,n}=\frac{2\mu\left(\,\omega_{\,\ell}-\gamma\,\right)}
{\left(\,\omega_{\,\ell}-\gamma\,\right)^2-\mu^2+2eB\,n}
$$
and, thereby
$$
\R\,\varkappa_{\,\ell,\,n\,,\,\pm}=
\rho_{\,\ell,\,n}\cos\alpha_{\,\ell,\,n}\mp\left(\,\omega_{\,\ell}-\gamma\,\right)\;,
$$
$$
\I\,\varkappa_{\,\ell,\,n\,,\,\pm}=
\rho_{\,\ell,\,n}\sin\alpha_{\,\ell,\,n}\mp\mu\;.
$$

It might be convenient to introduce another different parametrization for
the eigenspinors, which makes use of a pair of sequences of angles, namely,
$$
0<\theta_{\,\ell,\,n}^{\,\pm}<{\pi\over2}\,,\qquad
0<\Theta_{\,\ell,\,n}^{\,\pm}<2\pi\,,
\qquad\quad
\left(\,\ell\in{\mathbb Z}\,,\,n\in \mathbb N\,\right)\,.
$$
Thus, after calling
$$
\varkappa_{\,\ell,\,n\,,\,\pm}\equiv \tan\theta_{\,\ell,\,n}^{\,\pm}\
\exp\{\,i\,\Theta_{\,\ell,\,n}^{\,\pm}\,\}\;,
$$

we can also write

\beq
\Psi^{\,\pm}_{\ell,\,p,\,n}&=& \frac{\left(\,2eB\,\right)^\frac14}{\surd{\,4\pi\beta}}\
\exp\{\,i\tau\,\omega_{\,\ell} + i\,p\,y\}\no
&\times& \left\lgroup\begin{array}{c}
\pm\,i\ \phi_{\,n}(z)\ \cos\theta_{\,\ell,\,n}^{\,\pm}\\
\exp\{\,i\,\Theta_{\,\ell,\,n}^{\,\pm}\,\}\
\phi_{\,n-1}(z)\ \sin\theta_{\,\ell,\,n}^{\,\pm}
\\
\end{array}\right\rgroup\;.
\eeq
\end{enumerate}

Also in this case, the degeneracy {\it per} unit area is given by the Landau factor $\Delta$ in (\ref{landauB}).
We remark that, when going to the other non-equivalent representation, the symmetric part of the spectrum remains the same, but the asymmetric part transforms according to $\lambda_\ell \rightarrow -\lambda_\ell$ or, equivalently, $v\rightarrow -v$.

\section{The effective action. Invariance under large gauge transformations in the case of a complex chemical potential}\label{sect3}

It is important to note that, as can be seen from (\ref{asymspec}) and (\ref{symspec}), the whole spectrum is, in any of the two representations of the Clifford algebra, invariant under the so called {\em large} gauge transformations \cite{dunne,deser}
$$
\vu\ \rightarrow\ \vu^{\,\prime}=\vu + \frac{2\pi i\jmath}{\beta}\,,
\qquad\quad
\psi\,({\bf x})\ \rightarrow\ \psi^{\,\prime}\,({\bf x})=\psi\,({\bf x})\
\e^{\,i\tau\frac{2\pi\,\jmath}{\beta}}\,,
\qquad\left(\,\jmath\in\mathbb Z\,\right)
$$
where ${\bf x}=(x,y,\tau)\,,$ which are precisely those Euclidean gauge transformations leaving
untouched the antiperiodic boundary conditions satisfied by the
spinor Euclidean wave functions and, in consequence, the Fermi-Dirac statistics.
In other words, the discrete imaginary shift of the
chemical potential
$\beta v\ \rightarrow\ \beta v + 2\pi i\jmath\ (\,\jmath\in\mathbb Z\,)$
is a symmetry transformation of the theory.
On the contrary, any other imaginary shift of the chemical potential cannot be compensated by a
phase tranformation of the Euclidean spinor wavefunction, for the latter one would spoil the antiperiodic boundary conditions
that the Euclidean spinor field must satisfy in order to have the right statistics.

\bigskip

The Euclidean effective action for a sample of area
$\Omega\,,$ as defined in terms
of the zeta function regularization technique \cite{dowker}, is given by
\beq
\Gamma_{\,\rm eff} &=& \ln{Z}\equiv
\ln{\rm det}\Big(\ell_0\left(\,\Dirac\,\right)_{\,\rm Eucl}\Big)\nn\\
&\equiv& \Big[\,-\,\frac{d}{ds}\,
\zeta\left(\,s,\Dirac\ \ell_0\,\right)\,\Big]_{{s\,=\,0}}\no
&=& -\;\zeta^{\,\prime}\left(\,0,\Dirac\ \ell_0\,\right)\no
&=& {\Omega\Delta}\,[\,A\,(a)+S(a,b)\,]\;.
\eeq

In the last equation, we used the dimensionless reduced variables
$$
a \equiv  \frac{\beta\,\vu}{2\pi}\,,\qquad\quad
b \equiv {\textstyle\frac12}\,\beta\sqrt{2eB}\;,\nn
$$
the arbitrary length scale $\ell_0$ being introduced to render the
argument of the zeta function dimensionless. It is important to stress that physically meaningful results must be independent from $\ell_0$.
Here $A$ and $S$ denote the dimensionless reduced contributions to the full
effective action that originate respectively from the asymmetric and from the symmetric
part of the spectrum of the Euclidean Dirac operator.

In reference \cite{jphysa07}, we presented a detailed calculation of the
effective action in the case of a purely real chemical potential ($v=\mu$), and showed that different selections of the cut in the complex plane of the eigenvalues of (\ref{diracB}) during the evaluation of the determinant lead to predictions for the Hall conductivity which reproduce the behavior measured for mono- and bilayer samples.
In this section, we will concentrate on the contribution to the effective action due to the asymmetric
portion of the spectrum in the case of a complex chemical potential-- a survey of this calculation appears in \cite{leipzig}.

The contribution to
the effective action coming from the symmetric part
of the spectrum does not suffer from regularization
ambiguities. It is given by
$$
S(a,b)=2b\,\zeta_R\left(-{\textstyle\frac{1}{2}}\right)
+\sum_{n=1}^{\infty} \ln\left\{\left(1+z\,
\e^{-\beta{\ve}_n}\right) \left(1+z^{-1}\e^{-\beta
{\ve}_n}\right)\right\}\;,\nn
$$
where $z=\e^{\,2\pi a}$, $\ve_n\equiv\sqrt{2eB\,n}\,.$ and $\zeta_R$ is Riemann's zeta function. Notice that this
expression is invariant under the discrete symmetry
$v\,\to\,-\,v$, which is a symmetry of this portion of the spectrum.

On the other hand, the contribution to the effective action arising from the asymmetric part
of the spectrum is given by
$$
A(a)=-\, \zeta^{\,\prime}_{\,A}\left(\,0,a\,\right)\;,
$$
with
\beq
\fl\qquad
\zeta_{\,A}\,(\,s\,,\,a\,) &=&
\left(\frac{2\pi\ell_0}{\beta}\right)^{-s}\sum_{\ell\,=\,0}^\infty\left[\,
\left(\ell+\frac12+i\,\frac{\beta\,\vu}{2\pi}\right)^{-\,s}+
\left(-\,\ell-\frac12+i\,\frac{\beta\,\vu}{2\pi}\right)^{-\,s}\,\right]\no
&=& \left(\frac{2\pi\ell_0}{\beta}\right)^{-s}\sum_{\ell\,=\,0}^\infty
\left[\,(\,\ell+q_+\,)^{-\,s}+\e^{\,\mp\,i\pi s}(\,\ell+q_-\,)^{-\,s}\,\right]\;.
\label{ZetaHurwitz}
\eeq
The right hand side involves the sum of two series representations
of Hurwitz zeta functions, $\zeta(s,q)\,,$ which are well defined iff
$$
\R\,s>1\,,\qquad\quad
\beta\gamma\ \not=\ \pm\,\pi\,(2n-1)\,,\qquad (\,n\in\mathbb N\,)\,.
$$
In (\ref{ZetaHurwitz}), we have introduced the complex quantities
$$
q_{\,\pm}\ \equiv\ \frac12 \pm \frac{i\beta}{2\pi}\;(\,\mu+i\,\gamma\,)\,,
\qquad\quad
-\pi<\beta\gamma<\pi\;.
$$

The regularization ambiguity, which gives rise to a phase of the determinant \cite{ecz}, has its roots in the factor $e^{\,\mp\,i\pi s}$ or equivalently, in the selection of the cut of the complex power in the plane of the eigenvalues of (\ref{diracB}). Selecting one or the other sign (cut in the lower(upper) complex plane), irrespective of the sign of $\mu$, we get
\beq
\zeta_{\,A}\,(\,s\,,\,a\,) =
\left(\frac{2\pi\ell_0 }{\beta}\,\right)^{-\,s}
\Big\{\zeta\left(\,s,q_+\,\right)
+ \e^{\,\mp\,i\pi s}\,\zeta\left(\,s,q_-\,\right)\Big\}\;.
\nn
\eeq
For the contribution to the effective action we obtain, after evaluating the derivative,
\beq
A\,(a) &=&
\ln{2\pi}
-\,\ln\,{\Gamma(q_+)\,\Gamma(q_-)}
\pm{\rm i}\,\pi\left(\,{\textstyle\frac12}-q_-\,\right)\no
&=& \ln{2\pi}
-\,\ln\,{\Gamma\left({\textstyle\frac12}+{\rm i}a\right)\,
\Gamma\left({\textstyle\frac12}-{\rm i}a\right)}
\mp \pi a\;,
\eeq
$$
{\rm for}\qquad\quad
-\pi<\beta\,\gamma<\pi\,,\qquad\mu\in{\mathbb R}\;.
$$
The previous result can be also be cast in the form
\beq
A(a) &=&
\ln\,2\cosh\,\pi a \mp \pi a \no
&=&\ln\left(\,\e^{\,\pi a}+\e^{-\,\pi a}\,\right)
\mp \pi a\no
&=& \ln\left[\,\e^{\,{\textstyle\frac12}\,\beta(\mu+i\gamma)}+\e^{-\,\frac12\beta(\mu+i\gamma)}\,\right]
\mp \frac12\,\beta\,(\,\mu+i\gamma\,)\;.
\eeq

\bigskip
So, in our case, a well known sign
ambiguity \cite{ecz,deser} appears when evaluating the determinant of the Dirac operator,
as it happens whenever this last operator is not the square root of a second order operator of the Laplace type.
Note that, if the so called ``phase of the determinant" were chosen with the
same sign for all values of $\mu$,
then the contribution of each representation to the effective action would not
be an even function of $\mu$ for $\gamma=0$.
As a consequence, the contribution to the charge due to each Dirac point would
not change sign as $\mu \rightarrow -\mu\,.$

Now, it is always possible and convenient (although not mandatory)
to relate the two opposite phase prescriptions with the sign of the (real part of)
the chemical potential.

Turning back to eq.~(\ref{ZetaHurwitz}), if we cut the complex plane of the eigenvalues
$\lambda_{\,\ell}^{\,\pm}=(\,\ell+q_{\,\pm}\,)\quad(\,\ell=0,1,2,\ldots\,)$
along the upper half plane for positive values of $\mu$ and along the lower half plane for negative values of $\mu$, for example, we obtain what we will call the standard prescription,
and will identify through the index $\kappa=-1$. In this case we have
\beq
A(a)-\ln\,2\cosh\,\pi a = \left\{\begin{array}{cc}
-\,\pi\,a\,, & {\rm for}\ \mu>0\\
+\,\pi\,a\,, & {\rm for}\ \mu<0
\end{array}\right.\qquad (\,\kappa=-1\,)\,.
\eeq
We shall call the  non-standard prescription, $\kappa=+1$,
the opposite one,
\beq
A(a)-\ln\,2\cosh\,\pi a = \left\{\begin{array}{cc}
+\,\pi\,a\,, & {\rm for}\ \mu>0\\
-\,\pi\,a\,, & {\rm for}\ \mu<0
\end{array}\right.\qquad (\,\kappa=+1\,)\,.
\eeq
In so doing, for $\gamma=0$, we
recover our previous result \cite{jphysa07}, viz.,
\beq
{1\over\Omega}\,\Gamma_{\,\rm eff} &=&
\Delta\,\Big\{{\textstyle\frac12}\,\kappa\,\beta\,|\,\mu\,|+\ln\,2\cosh\,{\beta\mu\over2}
+ 2b\,\zeta\left(-{\textstyle\frac12}\right)\Big\}\no
&+& \Delta\,\sum_{n=1}^{\infty} \ln\left\{\left(1+z\,
\e^{-\beta{\ve}_n}\right) \left(1+z^{-1}\e^{-\beta
{\ve}_n}\right)\right\}\;.
\eeq

These two criteria lead, for a complex chemical potential such that $-\pi<\beta\,\gamma<\pi\,,$ to a contribution of the asymmetric part of the spectrum to the effective action, which is given by
\beq
A(a)&=& \ln\left(\,\e^{\,\pi a}+\e^{-\,\pi a}\,\right) + \pi a\kappa\,{\rm sgn}(\mu)\no
&=&\ln\left[\,\e^{\,\frac12\beta(\mu+i\gamma)}
+ \e^{-\,\frac12\beta(\mu+i\gamma)}\,\right]
+ {\textstyle\frac12}\,\kappa\,\beta\,(\,\mu+i\gamma\,)\,{\rm sgn}(\mu)\;.
\label{aeff}
\eeq
This contribution turns out to be invariant under the generalized conjugation symmetry
$a\to-\,a$.

One can take the limits $\beta\gamma\,\to\,\pm\,\pi$
in the last expression and write
$$
\lim_{\beta\gamma\,\to\,\pi^-}\,A(a)\ =\
\ln\,2\sinh\,{\beta\,|\,\mu\,|\over2}
+{\textstyle\frac12}\,\kappa\,\beta\,|\,\mu\,|
+{\pi i\over2}\,(\,1+\kappa\,)\,{\rm sgn}(\mu)\;,
$$
$$
\lim_{\beta\gamma\,\to\,-\,\pi^+}\,A(a)\ =\
\ln\,2\sinh\,{\beta\,|\,\mu\,|\over2}
+{\textstyle\frac12}\,\kappa\,\beta\,|\,\mu\,|
-{\pi i\over2}\,(\,1+\kappa\,)\,{\rm sgn}(\mu)\;,
$$
and therefore
$$
\left[\,\lim_{\beta\gamma\,\to\,\pi^-}\
-\ \lim_{\beta\gamma\,\to\,-\,\pi^+}\,\right] A(a)\
=\ \left\lbrace
\begin{array}{cc}
0 & {\rm for}\ \kappa=-1\\
2\pi i\,{\rm sgn}(\,\mu) & {\rm for}\ \kappa=+1
\end{array}\right.\;.
$$

However, the exact calculation can be performed at both extreme points of the interval, and both results coincide, as dictated by the invariance of the spectrum under large gauge transformations, which amounts to $\beta\gamma\,\to\,\beta\gamma+2\pi$. The zeta function regularization, which is based solely on the properties of the spectrum, respects this invariance for all values of $\beta\gamma$. In a zeta function regularization approach, not only the partition function is invariant under these transformations: it is the effective action itself that remains invariant. So, we can limit ourselves to give the result for the contribution to the effective action of the asymmetric part of the spectrum, for $-\pi<\beta\,\gamma\leq\pi\,,$
\bigskip

\beq
\Omega^{-1}\,\Gamma_{\,\rm eff} &=&
\,\Delta\,[\,A\,(a)+S(a,b)\,]\no
&=& {eB\over 2\pi}\left\lbrace
\ln\left(\,\e^{\,\pi a}+\e^{-\,\pi a}\,\right) + \pi a\kappa\,{\rm sgn}(\mu)+2b\,\zeta\left(-{\textstyle\frac12}\right)\right.\no
&+& \left.
\sum_{n=1}^{\infty} \ln\left[\,\left(1+z\,
\e^{-\beta{\ve}_n}\right) \left(1+z^{-1}\e^{-\beta
{\ve}_n}\right)\,\right]\right\rbrace\,.
\label{nose}\eeq

The result for any other value can be obtained by using
$$
\Gamma_{\,\rm eff}\,(\beta\mu+i\beta\gamma)=\Gamma_{\,\rm eff}\,(\beta\mu+2\pi
i\jmath+i\beta\gamma)\;,
\qquad\quad (\,\jmath\in{\mathbb Z}\,)\,.
$$
In fact, for any of the two selections of $\kappa$, the whole effective action, {\it per} unit degree of freedom is invariant under large gauge transformations. As explained in \cite{leipzig}, this amounts to an invariance under rotations of angle $4\pi$ around
an axis perpendicular to the plane of the sample in Euclidean space. We discuss this point in the next section.

\section{Zero temperature limit}\label{sect4}

Owing to the fact that the eigenfunctions corresponding to the asymmetric part of the spectrum
are eigenstates of $\sigma_3$ with eigenvalue $+1\,,$
one can equivalently write gauge transformations in the form
$$
\Psi_{\,\ell,\,p}\,(x,y,\tau) \rightarrow
\exp\left\{{\textstyle\frac12}\,{\rm i}\,
\sigma_3\,(2\gamma)\,\tau\right\}\,\Psi_{\,\ell,\,p}\,(x,y,\tau)\,.
$$
This last expression shows that, as the Euclidean time $\tau$ grows from $0$ to $\beta$,
spinors are rotated by $2\beta\gamma$, since $\frac12\,{\sigma_3}$
is the generator of rotations in the $Oxy-$ plane. In particular, $\beta\gamma=2\pi$ corresponds to a rotation of angle $4\pi$ around
the magnetic field. Such a rotation must be an invariance for spinors,
which provides one interpretation of the invariance under large
gauge transformations.

On the other hand, $\beta\gamma={\pi}$ corresponds to a rotation of angle
$2\pi\,.$ As can be seen from (\ref{nose}), at finite temperature, such a transformation turns the statistics from Fermi-Dirac
to Bose-Einstein,
for any of the two selections of phases in one of the two inequivalent representations (for a discussion about Bose-Fermi transmutation and relativistic "anyons" see, for instance, \cite{forte}.
For $\kappa=+1$, it also gives rise to an overall phase of
$\pi$ per unit degeneracy in the partition function, which can be recognized as
a Berry's phase. Such a phase is the contribution which survives
in the low temperature limit. In fact, in this limit one has from equation (\ref{aeff}),

\beq
\Omega\,\Delta\,A(a) =
{\textstyle\frac12}\,(\,\kappa+1\,)\,\Omega\Delta\,\beta\,(\,\mu+{\rm i}\,\gamma\,)\,{\rm sgn}(\mu)\,,
\qquad\quad \left(\,\beta\to\infty\,\right)\;.
\eeq

Thus, in the zero temperature limit, the effective action vanishes
for the standard phase prescription ($\kappa=-1$), so that the partition function
is rotationally invariant. For $\kappa=+1$, the imaginary term in $\Omega\,\Delta\,A(a)$
(i.e., the one proportional to $\gamma$) is the Abelian Chern-Simons term
\[{\rm i}\,\beta\,\gamma\,\Omega\,\frac{eB}{2\pi}\ {\rm sgn}(\mu)
={\mathcal A_{\,\rm CS}}
=\frac{\beta \Omega}{2}\,\kappa_{\,\rm CS}\ \frac{ie^2}{2\pi}\
\varepsilon^{\,\mu\nu\rho}\,F_{\,\mu\nu}\,A_{\,\rho}\;,\]
where $\kappa_{\,\rm CS}$ denotes the Chern-Simons coefficient, a pure
number of a topological origin
(remember that i$eA_3=\mu+{\rm i}\gamma$).
Our calculation clearly shows that the latter one  is invariant, as expected, under
large gauge transformations, which represent rotations by integer multiples of $4\pi\,.$
On the contrary, under rotations of angle $2\pi$ ($\beta\gamma=\pi$)
the partition function is multiplied by (when taking into account the two flavors)
$$
Z(\beta\gamma=\pi) \rightarrow Z\ \e^{\,\pm 2\pi i\,\Omega\Delta}\;.
$$

Hence, the requirement of  the partition function, in each representation, being invariant under
rotations of angle $2\pi$, for an area $\Omega$, entails a quantization condition
on the coefficient of the Chern-Simons term or, equivalently,
on the reduced flux of the magnetic field, i.e.,
$\Omega \Delta=N\in\mathbb N$. Interestingly enough, this is precisely
the condition for physical states to transform according to the one dimensional ray
representations of the magnetic translation group \cite{magtrans}.

Note also that the geometric phase appearing for $\kappa=+1$ leads, in the zero temperature limit, to a Berry's phase which equals $\pi$ for each degree of freedom. A similar calculation in the other irreducible representation of the Clifford algebra would give exactly the same results, if the same criterion were used to fix the phase of the determinant. As stressed in \cite{jphysa07}, the right behavior for the Hall conductivity in monolayer graphene appears when choosing opposite criteria in the two nonequivalent representations, which corresponds to a total Berry's phase of $\pi$ per unit degree of freedom. On the other hand, using the same criterium ($\kappa=+1$) in both representations, which corresponds to a Berry's phase of $2\pi$ per unit degree of freedom, correctly reproduces the Hall conductivity of bilayer graphene \cite{jphysa07}. For other discussions of Berry'a phases in mono- and bilayer graphene see, for instance, \cite{kope,falko,novo2}.

\section{Spinors in a constant electric field : Pair production at $T=0$}\label{sect5}

Consider now a graphene sample in the presence of a constant electrostatic field
pointing towards the positive
$Ox-$axis, i.e. $F^{10}=F_{01}=E_x=E>0\,.$ In this section, we will choose a
gauge which leads to non-stationary sets of solutions.
Throughout this section we will present a calculation, for the {\em 2+1}-dimensional case,
which is quite similar to the one performed in \cite{nikishov} for the {\em 3+1}-dimensional Minkowski
space-time.
In particular,
to solve the Dirac equation in the present {\em 2+1}-dimensional massless case
it is convenient to employ the following representation for the
gamma matrices,
\beq
\gamma^0=\sigma_1\,,\qquad\gamma^1=i\sigma_2\,,\qquad\gamma^2=i\sigma_3\,.
\eeq
After setting
$(x^0,x^1,x^2)=(t,x,y)$
we get the massless Dirac operator
in the gauge \cite{nikishov}
$A^\mu=(0,-Et,0)$, i.e.,
\beq
\Dirac = i\partial_t\,\sigma_1+(i\partial_x+eEt)\,i\sigma_2 - \partial_y\,\sigma_3\;,
\label{masslessdirac1}
\eeq
which is explicitly time dependent.
In order to obtain the solutions in this gauge,
it is convenient to introduce the partial Fourier transforms
\beq
\Psi(t,x,y)={1\over2\pi}\int_{-\infty}^{\infty}{dp}\int_{-\infty}^{\infty}dk\
\e^{ipx+iky}\,\widetilde \Psi(t,p,k)\,,
\label{fourier}
\eeq
with ${\bf p}=(p,k)$ as well as the dimensionless quantities
\beq
\xi &\equiv& (\,p-eEt\,)/\sqrt{eE}\,,
\qquad
\lambda\,\equiv\,{k^2}/{eE}\,.
\eeq

If we write
\beq
\Psi(t,x,y)=\left\lgroup\begin{array}{c}
\varphi(t,x,y)\\
\chi(t,x,y)\end{array}\right\rgroup\,,\qquad
\widetilde \Psi(t,p,k)=\left\lgroup\begin{array}{c}
\widetilde\varphi(t,p,k)\\
\widetilde\chi(t,p,k)\end{array}\right\rgroup\,,
\eeq
we obtain the coupled differential equations
\beq
\left\lbrace\begin{array}{c}
ik\,\widetilde\varphi + \sqrt{eE}\,(id_\xi+\xi)\,\widetilde\chi=0\,,\\
\sqrt{eE}\,(-id_\xi+\xi)\,\widetilde\varphi + ik\,\widetilde\chi=0\,.
\end{array}\right.
\eeq
Then, we can write
\beq
\widetilde\chi=(\,i/\sqrt\lambda)\,(-id_\xi+\xi)\,\widetilde\varphi\,,\qquad\quad\quad(\,\lambda\not=0\,)\\
(d_\xi^{\,2}+\xi^2+\lambda+i)\,\widetilde\varphi=0\;.\\
(id_\xi+\xi)\,\widetilde\chi=0\,,\qquad(-id_\xi+\xi)\,\widetilde\varphi=0\,,\qquad\quad(\,\lambda=0\,)\,.
\label{equaphi1}
\eeq

The general solutions of eq.~(\ref{equaphi1}) can be expressed in terms of two different sets of linearly independent solutions,
involving parabolic cylinder functions \cite{gradshteyn}.
\beq
\widetilde\varphi(\xi,\lambda) =
A_{\,\rm in}\,D_{-i\lambda/2}\,[\,+\,(1+i)\,\xi\,] +
B_{\,\rm in}^{\,\ast}\,D_{\,i\lambda/2-1}\,[\,+\,(1-i)\,\xi\,]\,,
\eeq
where $A_{\,\rm in}$ and $B_{\,\rm in}$ are complex constants.
From the recursive relations
\beq
(-id_\xi+\xi)\,D_{-i\lambda/2}\,[\,\pm\,(1+i)\,\xi\,] =
\mp\,{\lambda\over2}\,(1+i)\,D_{-i\lambda/2-1}\,[\,\pm\,(1+i)\,\xi\,]\,,
\no
(-id_\xi+\xi)\,D_{\,i\lambda/2-1}\,[\,\pm\,(1-i)\,\xi\,] =
\pm\,(1+i)\,D_{\,i\lambda/2}\,[\,\pm\,(1-i)\,\xi\,]\,,
\label{recursion}\eeq
we get the normalized spinors,
which are solutions of the original massless Dirac equation and read
\beq
_{-}\Psi_{\bf p}\,(t,{\bf x}) &=& {1\over4\pi}\,
\exp\left\{i\,{\bf p}\cdot{\bf x} - {\pi\lambda\over8}\right\}\no
&\times& \left\lgroup\begin{array}{c}
2D_{-i\lambda/2}\,[\,(1+i)\,\xi\,]\\
+\,(1-i)\sqrt\lambda\,D_{-i\lambda/2-1}\,[\,(1+i)\,\xi\,]\end{array}\right\rgroup\,,\\
_{+}\Psi_{\bf p}\,(t,{\bf x}) &=& {1\over4\pi}\,
\exp\left\{i\,{\bf p}\cdot{\bf x} - {\pi\lambda\over8}\right\}\no
&\times& \left\lgroup\begin{array}{c}
-(1+i)\sqrt\lambda\,D_{\,i\lambda/2-1}\,[\,(1-i)\,\xi\,]\\
2D_{\,i\lambda/2}\,[\,(1-i)\,\xi\,]\end{array}\right\rgroup\,.
\eeq
An alternative set of linearly independent solutions can be selected, that is
\beq
\fl\qquad
\widetilde\varphi(\xi,\lambda) =
A_{\,\rm out}\,D_{-i\lambda/2}\,[\,-\,(1+i)\,\xi\,] +
B_{\,\rm out}^{\,\ast}\,D_{\,i\lambda/2-1}\,[\,-\,(1-i)\,\xi\,]\,.
\eeq
Then, the recursive relations in equation (\ref{recursion}) give
\beq
^{+}\Psi_{\bf p}\,(t,{\bf x}) &=& {1\over4\pi}\,
\exp\left\{i\,{\bf p}\cdot{\bf x} - {\pi\lambda\over8}\right\}\no
&\times& \left\lgroup\begin{array}{c}
2D_{-i\lambda/2}\,[\,-(1+i)\,\xi\,]\\
-\,(1-i)\sqrt\lambda\,D_{-i\lambda/2-1}\,[\,-\,(1+i)\,\xi\,]\end{array}\right\rgroup\,,\\
^{-}\Psi_{\bf p}\,(t,{\bf x}) &=& {1\over4\pi}\,
\exp\left\{i\,{\bf p}\cdot{\bf x} - {\pi\lambda\over8}\right\}\no
&\times& \left\lgroup\begin{array}{c}
+(1+i)\sqrt\lambda\,D_{\,i\lambda/2-1}\,[\,-\,(1-i)\,\xi\,]\\
2D_{\,i\lambda/2}\,[\,-\,(1-i)\,\xi\,]\end{array}\right\rgroup\,.
\eeq
The above spinor solutions fulfill the orthonormality relations
\beq
\left({_\pm}\Psi_{\bf p}\,,\,_{\pm}\Psi_{\bf q}\right) =
\int{_\pm}\Psi_{\bf p}^{\,\dagger}\,(t,{\bf x})\,
 _{\pm}\Psi_{\bf q}\,(t,{\bf x})\ d{\bf x} = \delta({\bf p}-{\bf q})\;,
\label{on1}
\eeq
\beq
\left({^\pm}\Psi_{\bf p}\,,\,^{\pm}\Psi_{\bf q}\right) =
\int{^\pm}\Psi_{\bf p}^{\,\dagger}\,(t,{\bf x})\,
^{\pm}\Psi_{\bf q}\,(t,{\bf x})\ d{\bf x} = \delta({\bf p}-{\bf q})\;,
\label{on2}
\eeq
\beq
\left({_\pm}\Psi_{\bf p}\,,\,_{\mp}\Psi_{\bf q}\right)=0=
\left({^\pm}\Psi_{\bf p}\,,\,^{\mp}\Psi_{\bf q}\right)\;.
\label{on3}
\eeq

Note that for $\lambda=0=k\,,$ owing to $D_0(z)=\exp\{-z^2/4\}\,,$ both pairs of
solutions coincide and read
\beq
_{-}\Psi_p(t,{\bf x}) =\ ^{+}\Psi_p(t,{\bf x})&=& {1\over2\pi}\,
\exp\left\{i\,{px} -\,{i\over2}\,\xi^2(t)\right\}
\left\lgroup\begin{array}{c}
1\\
0\end{array}\right\rgroup\,,\\
_{+}\Psi_p(t,{\bf x}) =\ ^{-}\Psi_p(t,{\bf x}) &=& {1\over2\pi}\,
\exp\left\{i\,{px} +\,{i\over2}\,\xi^2(t)\right\}
\left\lgroup\begin{array}{c}
0\\
1\end{array}\right\rgroup\,.
\eeq
By looking at the asymptotic behavior of the parabolic cylinder functions, one finds

\beq
\fl
_{-}\Psi_{{\bf p}}\,(t,{\bf x}) \sim
{1\over2\pi}\,[\,2\xi^2(t)\,]^{-\,i\lambda/4}
\left\lgroup\begin{array}{c}
1\\
0
\end{array}\right\rgroup
\exp\left\{i\,{\bf p}\cdot{\bf x} - {i\over2}\,\xi^2(t)\right\}\,,
\qquad(\,t\to -\infty\,) \\
\fl
_{+}\Psi_{{\bf p}}\,(t,{\bf x}) \sim
{1\over2\pi}\,[\,2\xi^2(t)\,]^{\,i\lambda/4}
\left\lgroup\begin{array}{c}
0\\
1
\end{array}\right\rgroup
\exp\left\{i\,{\bf p}\cdot{\bf x} + {i\over2}\,\xi^2(t)\right\}\,,
\qquad(\,t\to -\infty\,) \eeq\beq
\fl
^{+}\Psi_{{\bf p}\,}(t,{\bf x})\sim
{1\over2\pi}\,[\,2\xi^2(t)\,]^{-\,i\lambda/4}
\left\lgroup\begin{array}{c}
1\\
0
\end{array}\right\rgroup
\exp\left\{i\,{\bf p}\cdot{\bf x} - {i\over2}\,\xi^2(t)\right\}\,,
\qquad(\,t\to+\infty\,)
\\
\fl
^{-}\Psi_{{\bf p}}\,(t,{\bf x}) \sim
{1\over2\pi}\,[\,2\xi^2(t)\,]^{\,i\lambda/4}
\left\lgroup\begin{array}{c}
0\\
1
\end{array}\right\rgroup
\exp\left\{i\,{\bf p}\cdot{\bf x} + {i\over2}\,\xi^2(t)\right\}\,.
\qquad(\,t\to +\infty\,)
\eeq

\medskip\noindent
Hence, we understand $_{+}\Psi_{{\bf p}}$ as the wave function of
an incoming quasi-particle, while we associate the wave function $_{-}\Psi_{{\bf p}}$ to an
incoming anti-quasi-particle.
Conversely, we shall associate $^{+}\Psi_{{\bf p}}$
and $^{-}\Psi_{{\bf p}}$ to the corresponding outgoing wave functions \cite{nikishov}.

\medskip
The above sets of incoming and outgoing solutions are related
throughout a Bogoliubov-like unitary transformation, that is

\beq
\fl\qquad\quad
\left\lbrace
\begin{array}{c}
_{+}\Psi_{{\bf p}}(x) = c_1\,^{+}\Psi_{\bf p}(x) + c_2\;^{-}\Psi_{\bf p}(x)\\
_{-}\Psi_{{\bf p}}(x) = -c_{2}^{*}\;^{+}\Psi_{\bf p}(x) + c_1^{*}\;^{-}\Psi_{\bf p}(x)
\end{array}\right.\;,\qquad\quad
|c_1|^2+|c_2|^2=1\;.
\label{liubov}
\eeq
We have
\beq
\left(\,{^+}\Psi_{\bf p}\,,\,_{+}\Psi_{\bf q}\,\right)=
N_\lambda^{*}\,\delta({\bf p}-{\bf q})\,,\\
\left(\,{^-}\Psi_{\bf p}\,,\,_{-}\Psi_{\bf q}\,\right)=
N_\lambda\,\delta({\bf p}-{\bf q})\,,\\
\left(\,{^\pm}\Psi_{\bf p}\,,\,_{\mp}\Psi_{\bf q}\,\right)=
\exp\left\{-\,{\pi\lambda/2}\right\}\,\delta({\bf p}-{\bf q})\,,
\eeq
where
\beq
N_\lambda\  =\
\sqrt{{\lambda\over\pi}}\;\Gamma\left({i\lambda\over2}\right)\,\sinh\left({\pi\lambda\over2}\right)
\exp\left\{-\,{\pi\lambda\over4}+{3\pi i\over4}\right\}\,.\label{enelambda}
\eeq
with
$$
|\,N_\lambda\,|^2\ =\ 1-\exp\left\{-\,{\pi\lambda}\right\}\,.
$$
Then, using the above listed orthonormality relations (\ref{on1}), (\ref{on2}) and (\ref{on3}),
we immediately find
\beq
c_1=N_\lambda^{*}\,,\qquad\quad c_2=\exp\left\{-\,{\pi\lambda/2}\right\}=c_2^{\,*}\,.
\label{ces}\eeq
As a consequence, in the transition to the quantum field theory, we can write the most general operator solutions
of the Dirac equation as

\beq
\Psi_{\rm in}(x) &=& \int d{\bf p}\
\left[\,a_{\,\rm in}({\bf p})\;_{+}\Psi_{\bf p}(x) +\
b_{\,\rm in}^{\,\dagger}({\,\bf p})\;_{-}\Psi_{{\bf p}}(x)\,\right]\,,
\nonumber\\
\Psi_{\rm out}(x) &=& \int d{\bf p}\
\left[\,a_{\rm out}({\bf p})\;^{+}\Psi_{\bf p}(x)\ +\
b_{\rm out}^{\,\dagger}({\bf p})\;^{-}\Psi_{\bf p}(x)\,\right]\,,
\nonumber
\eeq
where each set of creation and destruction operators fulfills the
standard canonical anticommutation relations.
From the condition $\Psi_{\rm in}(x)=\Psi_{\rm out}(x)$ one has

\beq
a_{\rm out}({\bf p}) &=& c_1\,a_{\,\rm in}({\bf p})-c_2\,b_{\,\rm in}^{\,\dagger}({\,\bf p})\,,\\
b_{\,\rm out}^{\,\dagger}({\,\bf p}) &=& c_{2}\,a_{\,\rm in}({\bf p})+c_1^{*}\,b_{\,\rm in}^{\,\dagger}({\,\bf p})\,,
\label{bogo}
\eeq
so that
\beq
\langle 0\,{\,\bf p}\,{\rm out}|&=&\langle 0\,{\,\bf p}\,{\rm in}|\,b_{\,\rm
  in}({\,\bf p})\,b_{\,\rm out}^{\,\dagger}({\,\bf p}^\prime\,)\no
&=&\langle 0\,{\,\bf p}\,{\rm in}|\,a_{\rm in}({\,\bf p})\,a_{\,\rm out}^{\,\dagger}({\,\bf p}^\prime\,)\no
&=&\langle 0\,{\,\bf p}\,{\rm in}|
\Big(c_1^{*}\,\delta({\bf p}-{\bf p}^\prime\,)-c_2\,a_{\,\rm in}({\,\bf p}^\prime\,)\,b_{\,\rm in}({\,\bf p})\Big)\ .
\label{vacua}
\eeq

\medskip
Once we have the  properly defined asymptotic {\sl in} and {\sl out} spinor
states, which are explicitly gauge dependent,
we can naturally define and calculate the gauge invariant probabilities for the
processes of pair creation and  annihilation in the presence of the background
electrostatic field.
We start by obtaining the vacuum persistence amplitude at a given ${\bf p}$. From
equations (\ref{ces}) and (\ref{vacua}) we find

\beq
\langle 0\,{\bf p}\,{\rm out}|0\,{\,\bf q}\,{\rm in}\rangle &=& c_1^{*}\,\delta({\bf p}-{\bf q})\no
&=&\int d{\bf x}\
^{-}\Psi^{\dagger}_{{\bf p}}(x)\,_{-}\Psi_{{\bf q}}(x)=
N_\lambda\,\delta({\bf p}-{\bf q})\,,
\eeq
with $N_\lambda$ given by equation (\ref{enelambda}).

In turn, the probability of vacuum decay at a given ${\bf p}$ is
$$
w_{\lambda}=1-|\,N_\lambda\,|^2\ ,
$$
or, more explicitly,
\beq
w_{\lambda}=\exp\left\{-\,{\pi\lambda}\right\}\,,\qquad
\lambda=\frac{k^2}{eE}\;.
\label{omega}\eeq
It is clear that the latter one is also the probability
of the inverse process, {i.e.} the annihilation in the state ${\bf p}$ with transfer of energy to the external
field. As a matter of fact, we also have
\beq
\int d{\bf x}\ ^{+}\Psi^{\dagger}_{{\bf p}}\,(t,{\bf x})\,
_{-}\Psi_{{\bf q}}\,(t,{\bf x})
= \delta({\bf p}-{\bf q})\,\exp\left\{-\,\textstyle\frac12\,{\pi\lambda}\right\}\,.
\eeq

\bigskip\noindent
The total vacuum persistence amplitude can
be written, after recovering physical units, in the form -- see the recent up to date review
\cite{dunne2},
\beq
\langle\,{\rm out}\,0\,|\,0\,{\rm in}\,\rangle=\exp\left\{{\rm
  i\over\hbar}\,[\,\R\,\Gamma_{\rm eff}(E) +{\rm i}\,\I\,\Gamma_{\rm eff}(E)\,]\right\}\ ,\eeq
and the corresponding probability is
\beq |\langle\,{\rm out}\,0\,|\,0\,{\rm in}\,\rangle|^{\,2}=
\exp\left\{-\,\frac{\,2\Omega\,\mathcal T}{\hbar}\,\Im{\rm m}\,{\cal L}_{\rm eff}(E)\right\}\,,
\label{eff_Lagrange}
\eeq
where ${\cal L}_{\rm eff}(E)$ is the effective Lagrange density
in the presence of a background electrostatic field $E\,,$
whereas $\mathcal T$ is the total time, and $\Omega$ denotes
the area of the sample.
It turns out that $\Gamma_{\rm eff}(E)$ contains a real part that describes dispersive
effects like the vacuum birefringence, as well as an imaginary part that
concerns absortive effects like vacuum decay.
Now, according to the above derivation, the vacuum persistence probability,
can be obtained from

\beq
\fl
\left(\Omega\,\mathcal T\right)^{-1}\ln{|\langle\,{\rm out}\,0\,|\,0\,{\rm in}\,\rangle|^{\,2}}=\frac{e\,E}{4\pi^2 \hbar^2}\int_{-\infty}^{\infty}
{dk}\ln{\left(1-\exp\left\{-\,{\frac{\pi\,k^2\,\tilde c}{e\,E\,\hbar}}\right\}\right)}\,\eeq
where we have introduced the number of states between ${\bf p}$ and ${\bf p}+d{\bf p}$, which is $\frac{\Omega}{4\pi^2 \hbar^2}$, and we have used that $dp=e\,E\,dt$.

So, for
the imaginary part of the effective action, we have, after taking the degeneracy $f$ into account,
\beq
{2\over\hbar}\,\Im{\rm m}\,{\cal L}_{\rm eff}(E)
&=& -\,f\,\frac{e\,E}{4\pi^2\, \hbar^2}\,
P.V.\int_{-\infty}^\infty dk\
\ln\left(1-\exp\left\{-\,{\frac{\pi\,k^2\,\tilde c}{e\,E\,\hbar}}\right\}\right)
\nonumber\\
&=& f\,\frac{(e\,E)^\frac32}{4\pi^2{\hbar}^{\frac32}\,\tilde c^{\frac12}}\,\sum_{n=1}^\infty {1\over n^{\,3/2}}
= f\,\left[\,{\zeta(\textstyle\frac32)\over2\pi}\,\right]\frac{eE}{2\pi\hbar}\sqrt{{eE\over\hbar\tilde c}}\ ,
\label{imlagrangeff}
\eeq
where $\tilde c$ is the speed of light for the two dimensional graphene sample
and $f$ is the number of fermion species or flavours, which is four for graphene,
since the result is the same in both nonequivalent representations of the Clifford algebra.
Moreover, the symbol $P.V.$ stands for the Cauchy principal value prescription,
i.e., $P.V.\int_{-\infty}^\infty dk=2\lim_{\,\epsilon\rightarrow 0}\int_{\epsilon}^\infty dk\,,$
which is the natural even prescription for the integrand around $k=0$. For previous, related results, see \cite{eff,moroz}.

Note that the vacuum persistence probability
$\exp\{(-2/\hbar)\,\Omega\,\mathcal T\,\Im{\rm m}\,{\cal L}_{\rm eff}(E)\}$ is quite small, even at finite time and for a finite area,
due to the massless character of the quasi-particles in graphene. So, the probability of the complementary event
\[P= 1-\exp{\left(-\,{2\over\hbar}\,\Omega\,\mathcal T\,\Im{\rm m}\,{\cal L}_{\rm eff}(E)\right)},\]
is high and by no means approximatively given by the effective action as in the case of massive particles/antiparticles pairs. The vacuum persistence probability obtained by replacing (\ref{imlagrangeff}) into this last equation coincides with the zero-mass limit of the one reported in \cite{allor}.

Some authors \cite{itz}, starting with J. Schwinger \cite{schwinger}, call this quantity the probability of pair creation.
Others (see \cite{allor,gavrilov} and references therein), following A.I. Nikishov \cite{niki2}, prefer to characterize
the creation of pairs through the rate of pair production. Note that the rate of pair production was evaluated in the same number of dimensions, for the massive case, in \cite{allor}. In the zero mass limit, the result found in this last reference is given by the first term of the series in (\ref{imlagrangeff}). This is exactly the result one would get after integrating (\ref{omega}) over all impulses, when counting the number of states between ${\bf p}$ and ${\bf p}+d{\bf p}$ as before.

Apart from the way one characterizes vacuum decay, the important point is, indeed, that the probability of actually detecting vacuum decay effects is another appealing characteristic of graphene, at variance with the case of massive particles, in which vacuum decay is very hard to be unraveled (see \cite{dunne2}).

\section{Dirac field interacting with a constant electric background at finite temperature and density}\label{sect6}

Consider now the problem of a graphene sample in the presence of a constant electric field
pointing towards the positive
$Ox$-axis, i.e. $F^{10}=F_{01}=E_x=E>0\,,$ at finite temperature and density.

As done in section \ref{sect2}, here we will employ path integral methods and zeta function regularization. In this section,
at variance with what has been done in the previous one,
we choose a gauge which leads to  a stationary set of solutions.

To solve for the eigenvalues of the Dirac equation in three dimensional Euclidean space, with metric $(+,+,+)$,
it is convenient to employ the following representation for the Euclidean
gamma matrices,
\beq
\egamma_1\,=\,\sigma_1\,,\qquad\egamma_2\,=\,\sigma_3\,,\qquad\egamma_3\,=\,\sigma_2\,.
\eeq

After setting
$ix_0=\tau\,,\ x_\mu=(x_1,x_2,x_3)=(x,y,\tau)$
and taking into account that
$\bar A_\mu=(0,0,{\mathcal E}x+i\upsilon/e)\,,\ (e>0)$ with $A_0=i\bar A_3\,,\ E\,=\,i\,{\mathcal E}\,,$
we get the Euclidean Dirac operator
\beq
\Dirac_{\;\rm Eucl} = (i\partial_\tau-e{\mathcal E}x-i\upsilon)\,\sigma_2+i\partial_x\,\sigma_1+i\partial_y\,\sigma_3\,.
\label{masslessdirac}
\eeq

In order to evaluate the effective action at finite temperature, one must impose antiperiodic boundary conditions
on the eigenfunctions, as in section \ref{sect2}. In this gauge,
it is convenient to write the partial Fourier transform

\beq
\Psi(x,y,\tau)={1\over\surd(2\pi\beta)}
\sum_{\ell\,=\,-\,\infty}^{\infty}\exp\{i\tau\omega_\ell\}
\int_{-\infty}^{\infty}{dk}\
\e^{iky}\,\widetilde \Psi_\ell(x,k)\,,
\label{fourier2}
\eeq
where, as before,

$$
\omega_\ell = {2\pi\over\beta}\left(\ell+\frac12\right)\,,\qquad\quad
$$
denote the Matsubara angular frequencies.

Moreover, it turns out to be very convenient to introduce the dimensionless quantities

\beq
\xi_\ell \equiv (\,\lambda_\ell + e{\mathcal E}x\,)/\sqrt{e{\mathcal E}}\,,\qquad\quad
d_\ell \equiv {d\over d\xi_\ell}\qquad\quad(\,\ell\,\in\,{\mathbb Z\,})\,,
\eeq
with $\lambda_\ell=\omega_\ell+iv\,,$ so that the eigenvalue equation reads (note that,
here, $\lambda$ represents the eigenvalues of the Euclidean Dirac operator,
and is different from $\lambda$ in the previous section)

\beq
\sqrt{e{\mathcal E}}
\left\lgroup\begin{array}{cc}
-k/\sqrt{e{\mathcal E}} & id_\ell+i\xi_\ell\\
id_\ell-i\xi_\ell & k/\sqrt{e{\mathcal E}}
\end{array}\right\rgroup\,\widetilde\Psi_\ell\,(x,k)=\lambda\,\widetilde\Psi_\ell\,(x,k)\,.
\eeq

If we write once again

\beq
\widetilde\Psi_\ell\,(x,k)\ \equiv\ \left\lgroup\begin{array}{c}
\widetilde\varphi_\ell\,(x,k)\\
\widetilde\chi_\ell\,(x,k)\end{array}\right\rgroup\,,
\eeq
we are lead to the coupled differential equations

\beq
\left\lbrace\begin{array}{c}
\sqrt{e{\mathcal E}}\,(d_\ell+\xi_\ell)\,\widetilde\chi_\ell+i\,(\lambda+k)\,\widetilde\varphi_\ell = 0\;,\\
\sqrt{e{\mathcal E}}\,(d_\ell-\xi_\ell)\,\widetilde\varphi_\ell + i\,(\lambda-k)\,\widetilde\chi_\ell = 0\;.
\end{array}\right.
\eeq

After solving these equations,

\medskip\noindent
(i)\quad
When $\lambda=k\,,$ we obtain the following infinite set
of degenerate normalized improper eigenspinors : namely,

\beq
\Psi_{0,\,k,\,\ell}\,(x,y,\tau)&=&\left({e{\mathcal E}\over\pi}\right)^\frac14\,
\exp\{\,i\tau\omega_\ell+iky\}\,(\,2\pi\beta\,)^{-\frac12}\no
&\times&\left\lgroup
\begin{array}{c}
0\\
\exp\left\{-\frac12\,e{\mathcal E}\left(x+\lambda_\ell/e{\mathcal E}\right)^2\right\}
\end{array}\right\rgroup
\,,\no
&&\ell\,\in\,{\mathbb Z}\,,\quad k\in{\mathbb R}\,,
\eeq
which fulfill
\beq
\fl\qquad
\int_0^\beta d\tau\int_{-\infty}^\infty dx\int_{-\infty}^\infty dy\
\Psi^\dagger_{0,\,p,\,m}\,(x,y,\tau)\,\Psi_{0,\,k,\,\ell}\,(x,y,\tau)\,=
\,\delta_{\ell m}\,\delta(k-p)\;.
\eeq

Their degeneracy is
\beq
&&\sum_{\ell\,=\,-\infty}^{\infty}\Psi^\dagger_{0,\,k,\,\ell}\,(x,y,\tau)\,
\Psi_{0,\,k,\,\ell}\,(x,y,\tau)=\no
&&{1\over 2\pi\beta}\,\sqrt{{e{\mathcal E}\over\pi}}
\sum_{\ell\,=\,-\infty}^{\infty}
\exp\left\{-\,\frac{4\pi^2}{e{\mathcal E}\beta^2}\left(\ell+\frac12+z\right)^2\right\}\
=\ {e{\mathcal E}\over4\pi^2}\;,
\eeq
where we have suitably introduced the rescaled and dimensionless complex coordinate
$z=(e{\mathcal E}x+i\upsilon)\frac{\beta}{2\pi}\,,$ while
use has been made of the Euler-McLaurin
summation formula -- see for example equation (3.6.28) in ref.~\cite{abramowitz}.
Note that $\ell-$th term of the series is an even function of $\ell$. Thus, all the odd derivatives vanish at $\ell=0$.
Moreover, the the $\ell-$th term and all its derivatives vanish at infinity. As a consequence we come to the result
that the number of degenerate eigenspinors {\it per} unit Euclidean "time" and  unit x-length is provided by
$
\Delta_{\,{\mathcal E}}={e{\mathcal E}}/{4\pi^2}\,,
$
the degeneracy factor being the same for all the remaining modes.

\medskip\noindent
(ii)\quad
For $\lambda\neq k$ we can write
\beq
\widetilde\chi=
\frac{i\sqrt{e{\mathcal E}}}{\lambda-k}\,(d_\ell-\xi_\ell)\,\widetilde\varphi\,,\\
(d_\ell^{\,2}-\xi_\ell^2+\Lambda-1)\,\widetilde\varphi=0\,,\qquad\quad
\Lambda\equiv\frac{\lambda^2 -k^2}{e{\mathcal E}}\ .
\label{equaphi}
\eeq
As in section \ref{sect2}, the eigenvalue problem for each $\ell\in{\mathbb Z}$ is the one
of a linear harmonic oscillator, so that the spectrum takes the symmetric form
$$
\lambda_{\,k\,,\,n}\,=\,\pm\,\sqrt{k^2+2e{\mathcal E}n}\,,\qquad\quad
n=1,\ldots,\infty\,,\quad k\in{\mathbb R}\,,\qquad\quad
(\,\forall\,\ell\,\in\,{\mathbb Z}\,)\,.
$$

The somewhat surprising conclusion we can draw from the above analysis is that neither the
eigenvalues nor the degeneracy depend at all upon the temperature and/or the chemical potential.
Hence, once we have at hand the spectrum of the Euclidean Dirac operator and its overall degeneracy,
which is the same for all eigenvalues, we can turn to the evaluation of the Euclidean effective
action, within the zeta function approach. In this case we have

\beq
\fl
\zeta\left(\,s,{\Dirac}\,\ell_y\,\right)=
\Delta_{\;\!{\mathcal E}}\,\beta\,\ell_x\,\ell_y\int_{-\infty}^{\infty}\,dk\left[\,
(\ell_y k)^{-s} +
\frac{1+\e^{\,\mp\,\pi is}}{(2e{\mathcal E}{\ell_y}^{\,2})^{s/2}}
\sum_{n=1}^{\infty}\,\left(n+\frac{k^2}{2e{\mathcal E}}\right)^{-\,s/2}\,\right]\,.
\label{zeta1}
\eeq

Here, the length scale $\ell_y$ has been introduced to render the second argument in the zeta function dimensionless.

Now, eq.~(\ref{zeta1}) can be rewritten as

\beq
&&\zeta\left(\,s,{\Dirac}\,\ell_y\,\right)\ =\ \Delta_{\;\!{\mathcal E}}\,\Omega\,\beta\left\lbrace\,
\int_{-\infty}^{\infty}\,dk\ (\ell_y\, k)^{-s}\right.\ +
\label{zeta2}\\
&&\left.\frac{1+\e^{\,\mp\,\pi is}}{\Gamma({s}/{2})\left(2e{\mathcal E}{\ell_y}^2\right)^{{s}/{2}}}
\int_{-\infty}^{\infty}\,dk\int_{0}^{\infty}dt\;t^{{s}/{2}-1}\,
\sum_{n=1}^{\infty}\,\e^{-\,tn-t{k^2}/{2e{\mathcal E}}}\right\rbrace\,.
\nn
\label{nuevenueve}\eeq
For $\R s>1$ the two integrals in the second line can be interchanged and, if the integral in the first line is
understood in terms of the Cauchy principal value prescription,
according to our comment in section \ref{sect3}, we get

\beq
\fl
\zeta\left(\,s,{\Dirac}\,\ell_y\,\right)\ =\
\Delta_{\;\!{\mathcal E}}\,\Omega\beta\left(1+\e^{\,\mp\,\pi is}\right)\,{\ell_y}^{-s}\, \times \nn\\
\fl \left[\,\lim_{\epsilon \rightarrow 0}\frac{(\,\epsilon\,)^{1-s}}{{1-s}}+\frac{(s/2)\sqrt\pi}{\Gamma\left(1+{s}/{2}\right)}\,
\left(2e{\mathcal E}\right)^{\frac{1-s}{2}}\,\zeta\left(\frac{s-1}{2}\right)\,\Gamma\left(\frac{s-1}{2}\right)\,
\right]\,.
\label{zeta3}
\eeq

We finally obtain the Euclidean effective action per unit
volume in the form (note, however, that the contribution coming from the first term was defined here as
$-\lim_{\epsilon\,\rightarrow\,0}\,\left.
\int_{\epsilon}^{\infty}\,dk\,\frac{d}{ds}\right\rfloor_{s\,=\,0}\,(\ell_y\,\,k)^{-s}\,\,$,
the integrand being evaluated at $\R s>1$ and then extended to $s=0$ before taking the principal value limit)

\beq
{1\over\Omega\beta}\left[\,-\,\frac{d}{ds}\,\zeta\left(\,s,{\Dirac}\;\beta\,\right)\,\right]_{\,s\,=\,0}
&=&
-\Delta_{\;\!{\mathcal E}}\,\Gamma\left(-{\textstyle\frac12}\right)\zeta
\left(-{\textstyle\frac12}\right)\sqrt{2\pi e{\mathcal E}}\no
&=& -\,2\sqrt2\,\pi\,\Delta_{\;\!{\mathcal E}}\,
\zeta\left(-{\textstyle\frac12}\right)\sqrt{e{\mathcal E}}\,.\nn
\eeq
where (see \cite{gradshteyn})
\beq
2\sqrt2\,\zeta\left(-\,{\textstyle\frac12}\right) &=&
\frac{1}{\pi}\sum_{n=1}^\infty
n^{-3/2}\,\sin\left(-{\pi\over4}\right)\,=\,-\,{\zeta\left({\textstyle\frac32}\right)\over\pi\surd2}\,.
\nonumber
\eeq

As a consequence, we can recast the Euclidean effective Lagrangean as
\beq
{\mathcal L}_{\,\rm eff}^{\,\rm E}
=\,\frac{\,\zeta\left({\textstyle\frac32}\right)}{4\pi^2\surd\,2}\,(e{\mathcal E})^{3/2}\,,
\label{accion}
\eeq
so that, after turning back to the {\em 2+1} dimensional Minkowski space-time
by means of the replacement ${\mathcal E}=-\,{\rm i}E\,,$ we obtain the
effective Lagrange density
\beq
{\mathcal L}_{\,\rm eff}^{\,\rm
  M}(E)\,=\,\frac{\zeta\left({\textstyle\frac32}\right)}{8\pi^2}\,(eE)^{3/2}
[\,(1+{\rm i})\,]\,.
\eeq
It follows that, as in eq.~(\ref{eff_Lagrange}),
we end up with the correspondence

\beq
\fl\qquad
\langle\,{\rm out}\,0\,|\,0\,{\rm in}\,\rangle =
\exp\{{\rm i}\,\Omega\,{\mathcal T}{\mathcal L}_{\rm eff}^{\,\rm M}(E)\}=
\exp\left\{{\rm i}\,\Omega\,{\mathcal T}\,
\frac{\zeta\left({\textstyle\frac32}\right)}{8\pi^2}\,(eE)^{3/2}(1+{\rm i})\right\}\,,
\eeq
which leads, after recovering physical units, to eq.~(\ref{imlagrangeff}), i.e., the {\it 2+1} dimensional
counterpart of the celebrated Schwinger formula \cite{schwinger}
\beq
\I\,{\mathcal L}_{\rm eff}^{\,\rm M}(E)\,=\,f\;
\frac{\zeta\left({\textstyle\frac32}\right)(eE)^\frac32}{8\pi^2\hbar\surd{\hbar\tilde c}}
\,,
\eeq
where $\tilde c$ is the speed of light for the two dimensional graphene sample
and $f$ is the number of fermion species or flavours, which is four for graphene,
in perfect agreement with our previous result (\ref{imlagrangeff}).
Moreover we find that in the {\it 2+1} dimensional case the following remarkable
equalities hold true, viz.,
$$
\R\,{\mathcal L}_{\rm eff}^{\,\rm M}(E)\,=\,\I\,{\mathcal L}_{\rm eff}^{\,\rm M}(E)
\,=\,-\,{\mathcal L}_{\rm eff}^{\,\rm E}({\mathcal E})\,.
$$

\section{Minimal quantum conductivity of graphene from planar QED}
\label{sect7}

Of particular interest in the presence of a constant electric background is the mean current density or,
equivalently, the minimal conductivity of graphene.
In order to evaluate the mean value of such quantum current density, we go back to our calculation in the previous section,
this time introducing a Lagrange multiplier $\alpha$ through the non-trivial replacement $k\rightarrow (k+i\alpha)\,$.
The derivative of the effective action with respect to the external parameter $\alpha$
will then give us the required mean value of the quantum current density.
Actually, we can say that the control parameter $\alpha$ plays a very similar role
to the one played by the chemical potential in the evaluation
of the particle number or charge density.

If we consider the manifold to be compact in the $y$ direction, all we have to do is to turn back to equation (\ref{nuevenueve}),
where we have, after a shift in the transverse momentum integration variable $k\,,$
\beq
&&\zeta\left(\,s,{\Dirac}\,\ell_y\,; \alpha\,\right)\ =\
\Delta_{\;\!{\mathcal E}}\,\Omega\,\beta\,{\ell_y}^{\,-s}\left\lbrace\,
\int_{-\infty}^{\infty}\,dk\ (k+i\alpha)^{-s}\right.\ +
\label{zeta2bis}\\
&&\left.\frac{1+\e^{\,\mp\,\pi is}}{\Gamma({s}/{2})\left(2e{\mathcal E}\right)^{{s}/{2}}}
\int_{-\infty}^{\infty}\,dk\int_{0}^{\infty}dt\;t^{{s}/{2}-1}\,
\sum_{n=1}^{\infty}\,\e^{-\,tn-t{(k+i\alpha)^2}/{2e{\mathcal E}}}\right\rbrace\,.
\nn
\label{nuevenuevebis}
\eeq

Now, the shift in the second term is irrelevant, since the Gaussian function is an entire function. Then, the
$\alpha-$dependence is solely due to the first term and, consequently, the only contribution
to the minimal quantum current density will arise from the quantity
\beq
Z_{\;\!{\mathcal E}}(s\,;\alpha)&=&\Delta_{\;\!{\mathcal E}}\,\Omega\,\beta\,{\ell_y}^{-s}\,
\int_{-\infty}^{\infty}\,dk\ (k+i\alpha)^{-s}\no
&=&\Delta_{\;\!{\mathcal E}}\,\Omega\,\beta\,{\ell_y}^{-s}\,
\int_{0}^{\infty}\,dk\ \left[(k+i\alpha)^{-s}+(-1)^{-s}(k-i\alpha)^{-s}\right]\,.\nn
\eeq

At variance with the case of the previous section ($\alpha=0$), the zeta regularization of this term is now well defined, the only ambiguity being the selection of the phase, much as in the calculation of the charge
when the sample is subjected to a magnetic field. As in that case, one can make two selections of the phase.
Indeed,

\beq
\fl\qquad
Z_{\;\!{\mathcal E}}(s\,;\alpha)=\Delta_{\;\!{\mathcal E}}\,\Omega\,\beta\,{\ell_y}^{-s}\,
\int_{0}^{\infty}\,dk\ \left[(k+i\alpha)^{-s}+\e^{-\pi is\kappa\,{\rm sign}(\alpha)} (k-i\alpha)^{-s}\right]\,,
\eeq
where, as before, $\kappa=-1$ represents the "usual" selection of phase, whilst $\kappa=1$ the opposite
"unusual" one. Thus, for $\R\,s>1\,,$ we get

\beq
Z_{\;\!{\mathcal E}}(s\,;\alpha)=-\Delta_{\;\!{\mathcal E}}\,\Omega\,\frac{(i\alpha\beta)^{1-s}}{1-s}\,
\left[1-\e^{-2\pi is\kappa\,{\rm sign}(\alpha)}\right]\,.
\eeq
Performing the $s-$derivative and evaluating at $s=0$, we get for the $\alpha-$dependent part of
the Euclidean effective action,

\beq
\Gamma^{\,\rm E}_{\rm eff}(\alpha)= -2\pi\,\Delta_{\;\!{\mathcal E}}\,\Omega\,\beta\,\kappa\,|\alpha|
= -\frac{e{\mathcal E}}{2\pi}\,\beta\,\Omega\,\kappa\,|\alpha|\,.
\eeq
To the aim of recovering the minimal quantum current density it is necessary, first, to go back to the
{\it2+1} dimensional Minkowski space-time, i.e.,
\beq
\Gamma^{\,\rm M}_{\rm eff}(\alpha)=\,\frac{eE}{2\pi}\,{\mathcal T}\,\Omega\,\kappa\,|\alpha|\,.
\eeq
Now, it turns out that performing the derivative with respect to $\alpha\,,$ dividing by
$\Omega\,{\mathcal T}$ and multiplying by the elementary charge, $-e$, we have
\beq
\langle\,J_{\rm min}\,\rangle =-\,\frac{e^2 E}{2\pi}\kappa\,{\rm sign}(\alpha)\,,
\eeq

Note that, in the other representation, the result is the same if the same criterium is chosen
to define the phase of the determinant.
Thus, summing up the contributions from both representations,
contrary to the case of the Hall conductivity, with the same criterium,
multiplying by the two spins (flavors), and recovering the physical units,
we find for graphene
\beq
\langle\,J_{\rm min}\,\rangle = -\,\frac{4e^{2}}{h}\,E\,\kappa\,{\rm sign}(\alpha)\,.
\eeq
The quantization of the minimal quantum conductivity
\beq
\sigma_{\rm min} = -\,\frac{4e^{2}}{h}\,\kappa\,{\rm sign}(\alpha)
\eeq
in terms of the quantum unit of conductivity makes this prediction entirely different
from the results obtained, for instance, through the Kubo formula \cite{mincond,ludwig94}, where an extra factor of $\pi$ in the denominator appears. Moreover,
for $\kappa=-1$ and positive $\alpha$ one obtains exactly the result in \cite{geim}. It's interesting to note that a similar experimental value is found also for bilayer graphene \cite{novo2}.

As already stressed, this result is independent both from the temperature and the chemical potential.

\section{Conclusions}\label{conclusions}

In this paper we have carefully analyzed the response of a graphene sample
under the influence of homogeneous electric and magnetic fields.
Our present investigation has its roots in the well established
quantum field theoretic model of a free massless Dirac spinor in two space and
one time dimensions, i.e. the massless planar spinor quantum electrodynamics.
In spite of its simplicity, this quantum field theory does exhibit remarkable features. As a matter of fact,
in the presence of a uniform magnetic field at finite temperature and density,
a nice connection has been established, in this manuscript, among the phase prescription of the effective action,
the large gauge transformation invariance, Berry's phases and Chern-Simons topological terms,
besides the different forms for the Hall conductivity we had already elucidated in our previous paper
\cite{jphysa07}.

Moreover, in the presence of a constant electrostatic field,
a derivation of the imaginary part of the effective Lagrangian at zero temperature has been presented. Our result coincides with the zero mass limit of the one obtained, for instance, in \cite{allor}.

Interestingly enough, at finite temperature, the imaginary part of the effective Lagrangian turns out to be independent of
the temperature and density and coincides
with the massless planar limit of the celebrated Schwinger formula \cite{schwinger}. The corresponding probability of vacuum decay, for a realistic graphene sample, can be very high for laboratory fields, at variance with the common case
of massive charged electron positron pairs. In the case of $3+1$ dimensions, such independence was shown to hold, to the order of one loop, in reference \cite{dittrich}. Our result in $2+1$ is stronger: indeed, our calculations  show that the whole effective lagrangian is independent of the temperature and of the chemical potential, within the zeta function regularization scheme.

Furthermore, our treatment, based on the zeta function regularization technique,
allowed us to obtain the (finite) average value for the minimal quantum current density and, thus, a
minimal quantum conductivity for graphene,
which, at variance with most theoretical results \cite{mincond,ludwig94} agrees, for a particular selection of the phase of the determinant in both irreducible representations, with the one exhibited in \cite{geim}. The corresponding current is topological in origin, as is the behavior of the Hall conductivity (for general studies of topological currents in 2+1 QED see, for example, \cite{quique,moroz}). We hope
the present paper will be helpful in clarifying the so called "mystery of the missing pi" \cite{geim},
or its absence thereof \cite{rutgers}.

\ack{This work was partially supported by Universidad Nacional de La Plata
(Proyecto 11/X492) and CONICET (PIP 6160). P.G. and R.S. would like to thank
the Istituto Nazionale di Fisica Nucleare for grant I.S. PI13. E.M.S. thanks S. Gavrilov, T. Cohen and N. Protasov for useful comments and discussions.}

\appendix
\section*{Appendix : Parabolic cylinder functions}

The parabolic cylinder functions, of the special form we are interested in the present context,
can be defined, e.g., by the integral representation
{\bf 9.241} 1. p. 1092 of ref.~\cite{gradshteyn}
\beq
D_{-i\lambda/2}\,[\,\pm(1+i)\,\xi\,] &=&
{1\over \surd\pi}\,2^{-i\lambda/2+1/2}\,\e^{-\pi\lambda/4}\,\e^{i\xi^2/2}
\nonumber\\
&\times& \int_{-\infty}^\infty x^{-i\lambda/2}\,\e^{-2x^2\pm 2ix(1+i)\xi}\ dx\,,
\label{parcyldef}
\eeq
where $\lambda>0\,,\ \xi\in{\mathbb R}\,,\ {\rm arg}\,x^{-i\lambda/2}=\lambda/2$ for $x<0\,,$ so that
\beq
D^{\,*}_{-i\lambda/2}\,[\,\pm(1+i)\,\xi\,] &=&
{1\over \surd\pi}\,2^{\,i\lambda/2+1/2}\,\e^{-\pi\lambda/4}\,\e^{-i\xi^2/2}
\nonumber\\
&\times& \int_{-\infty}^\infty x^{i\lambda/2}\,\e^{-2x^2\mp 2ix(1-i)\xi}\ dx\,.
\eeq
After the change of variable $x\,\longmapsto\,-\,x$
\beq
D^{\,*}_{-i\lambda/2}\,[\,\pm(1+i)\,\xi\,] &=&
{1\over \surd\pi}\,2^{\,i\lambda/2+1/2}\,\e^{\pi\lambda/4}\,\e^{-i\xi^2/2}
\nonumber\\
&\times& \int_{-\infty}^\infty x^{i\lambda/2}\,\e^{-2x^2\pm 2ix(1-i)\xi}\ dx\,,
\eeq
we eventually come to the conjugation property
\beq
D^{\,*}_{-i\lambda/2}\,[\,\pm (1+i)\,\xi\,]\ =\
D_{\,i\lambda/2}\,[\,\pm (1-i)\,\xi\,]\,,
\label{parcylconj}
\eeq
as na\"\i vely expected.
The following special values appear in our calculations,
\beq
&& D_{\,\pm i\lambda/2}\,(0)=\pi^{-1/2}\,2^{\pm i\lambda/4}\,
\Gamma\left(\frac12\pm{i\lambda\over 4}\right)\,\cosh{\pi\lambda\over 4}\,,\\
&& D_{\,\pm i\lambda/2-1}\,(0)=\pm i\,\pi^{-1/2}\,2^{\pm i\lambda/4-1/2}\,
\Gamma\left(\pm{i\lambda\over 4}\right)\,\sinh{\pi\lambda\over 4}\,;\\
&& \pm\,{\lambda\over 2}\,|\,D_{\,\pm i\lambda/2-1}\,(0)\,|^2 =
\pm\,\sinh{\pi\lambda\over 4}\,,\\
&& |\,D_{\,\pm i\lambda/2}\,(0)\,|^2=\cosh{\pi\lambda\over 4}\,.
\eeq
The parabolic cylinder functions fulfill the recursion formulas
\beq
&& \frac{d}{dz}\,D_\nu(z) = -\,\frac12\,zD_\nu(z)+\nu\,D_{\nu-1}(z)\,,
\label{recursionlowering}\\
&& \frac{d}{dz}\,D_\nu(z) = \frac12\,zD_\nu(z) - D_{\nu+1}(z)
\label{recursionraising}\,.
\eeq
Consider the combination
\beq
D_+ &\equiv&
D_{-i\lambda/2}\,[\,(1+i)\,\xi\,]\,D_{\,i\lambda/2}\,[\,(1-i)\,\xi\,]
\nonumber\\
&+& {\lambda\over 2}\,D_{-i\lambda/2-1}\,[\,(1+i)\,\xi\,]\,
D_{\,i\lambda/2-1}\,[\,(1-i)\,\xi\,]\,.
\label{D_+}
\eeq
From the recursion formul\ae\  we get
\beq
&& 2\,{d\over d\xi}\,D_{-i\lambda/2}\,[\,(1+i)\,\xi\,]\,
   D_{\,i\lambda/2}\,[\,(1-i)\,\xi\,]\ =\nonumber\\
&& \lambda\,(1+i)\,D_{-i\lambda/2}\,[\,(1+i)\,\xi\,]\,
   D_{\,i\lambda/2-1}\,[\,(1-i)\,\xi\,]\ +\ {\rm c.c.}\nonumber\\
&& \lambda\,{d\over d\xi}\,D_{-i\lambda/2-1}\,[\,(1+i)\,\xi\,]\,
   D_{\,i\lambda/2-1}\,[\,(1-i)\,\xi\,]\ =\nonumber\\
&& -\,\lambda\,(1+i)\,D_{-i\lambda/2}\,[\,(1+i)\,\xi\,]\,
   D_{\,i\lambda/2-1}\,[\,(1-i)\,\xi\,]\ +\ {\rm c.c.}\,,\nonumber
\eeq
so that the above combination $D_+$ does not depend upon $\xi$
and from the conjugation property (\ref{parcylconj}) we can write
\beq
D_+ =\ |D_{-i\lambda/2}(0)|^{\,2} +
{\lambda\over 2}\,|D_{-i\lambda/2-1}(0)|^{\,2}
=\ \exp\{\pi\lambda/4\}\,.
\label{parcylnorma}
\eeq
Let us now consider the further combination
\beq
D_- &\equiv&
D_{-i\lambda/2}\,[\,(1+i)\,\xi\,]\,D_{\,i\lambda/2}\,[\,-(1-i)\,\xi\,]
\nonumber\\
&-& {\lambda\over 2}\,D_{-i\lambda/2-1}\,[\,(1+i)\,\xi\,]\,
D_{\,i\lambda/2-1}\,[\,-(1-i)\,\xi\,]\,.
\label{D_-}
\eeq
From the recursion formul\ae\  we get
\beq
&& 2\,{d\over d\xi}\,\{D_{\,i\lambda/2}\,[\,(i-1)\,\xi\,]\,
D_{\,-i\lambda/2}\,[\,(i+1)\,\xi\,]\}\ =\nonumber\\
&& \lambda\,(1-i)\,D_{\,i\lambda/2}\,[\,(i-1)\,\xi\,]\,
   D_{\,-i\lambda/2-1}\,[\,(i+1)\,\xi\,]\ -\nonumber\\
&& -\,\lambda\,(1+i)\,D_{\,-i\lambda/2}\,[\,(i+1)\,\xi\,]\,
   D_{\,i\lambda/2-1}\,[\,(i-1)\,\xi\,]\;\nonumber\\
&& \lambda\,{d\over d\xi}\,\{D_{-i\lambda/2-1}\,[\,(i+1)\,\xi\,]\,
D_{\,i\lambda/2-1}\,[\,(i-1)\,\xi\,]\}\ =\nonumber\\
&& \lambda\,(1-i)\,D_{\,i\lambda/2}\,[\,(i-1)\,\xi\,]\,
   D_{\,-i\lambda/2-1}\,[\,(i+1)\,\xi\,]\ -\nonumber\\
&& -\,\lambda\,(1+i)\,D_{\,-i\lambda/2}\,[\,(i+1)\,\xi\,]\,
   D_{\,i\lambda/2-1}\,[\,(i-1)\,\xi\,]\,,
\eeq
which leads to the conclusion that also the quantity $D_-$
is independent of $\xi$, and yields
\beq
D_- =\ |D_{-i\lambda/2}(0)|^{\,2} -
{\lambda\over 2}\,|D_{-i\lambda/2-1}(0)|^{\,2}
=\ \exp\{-\pi\lambda/4\}\,.
\label{parcylpair}
\eeq
The above important properties of the parabolic cylinder functions
can be summarized in the remarkable formula
\beq
D_\pm =\ |D_{\,i\lambda/2}\,(0)|^{\,2} \pm
{\lambda\over 2}\,|D_{\,i\lambda/2-1}(0)|^{\,2}
=\ \exp\{\pm\pi\lambda/4\}\,.
\label{parcylfunda}
\eeq
Consider the second order differential equations
\beq
\left({d^2\over d\xi^{2}}+\xi^{2}+\lambda\pm\,i\right)f_\pm(\xi,\lambda)=0\,.
\eeq
Two pairs of linearly independent solutions for the upper sign equation are
\beq
f_+^{\,(1)}(\pm\,\xi,\lambda) = D_{-i\lambda/2}\,\left(\pm\,\xi\sqrt2\,\e^{\,\pi i/4}\right)\\
f_+^{\,(2)}(\pm\,\xi,\lambda) = D_{\,i\lambda/2-1}\,\left(\pm\,\xi\sqrt2\,\e^{\,-\,\pi i/4}\right)
\eeq
while two couples of linearly independent solutions for the lower sign equation are
\beq
f_-^{\,(1)}(\pm\,\xi,\lambda)=D_{-i\lambda/2-1}\,\left(\pm\,\xi\sqrt2\,\e^{\,\pi i/4}\right)\\
f_-^{\,(2)}(\pm\,\xi,\lambda)=D_{\,i\lambda/2}\,\left(\pm\,\xi\sqrt2\,\e^{\,-\,\pi i/4}\right)
\eeq
To the aim of verifying linear independence we have to compute the wronskian.
Let us first calculate derivatives by means of the recursion formul\ae\
(\ref{recursionlowering}) and (\ref{recursionraising}) that yield
\beq
\frac{d}{d\xi}\,D_{-i\lambda/2}\,\left(\pm\,\xi\sqrt2\,\e^{\,\pi i/4}\right) =\no
-\,i\,\xi\,D_{-i\lambda/2}\,\left(\pm\,\xi\sqrt2\,\e^{\,\pi i/4}\right)
\mp\,{\lambda\over\surd\,2}\,\e^{\,3\pi i/4}\,D_{-i\lambda/2-1}\,\left(\pm\,\xi\sqrt2\,\e^{\,\pi i/4}\right)\nn
\eeq
and thereby
\beq
W\left[\,f_+^{\,(1)}(\xi,\lambda)\,,\,f_+^{\,(1)}(-\,\xi,\lambda)\,\right]
&=& {1+i\over\surd\,\pi}\,\Gamma\left(-\,{i\lambda\over2}\right)\,\sinh\left({\pi\lambda\over2}\right)
\eeq
On the other side we readily find
\beq
W\left[\,f_+^{\,(1)}(\pm\,\xi,\lambda)\,,\,f_+^{\,(2)}(\pm\,\xi,\lambda)\,\right]\ =\
\mp\,(1-i)\,\exp\{\,\pi\lambda/4\}
\eeq
and analogous relationships for the other solutions.

\bigskip
In order to understand the physical meaning of the solutions
of the wave field equations,
we have to analyze the leading asymptotic behavior of
the parabolic cylinder functions.
Then
from eq.~{\bf 9.246} 1. p. 1093 of
ref.~\cite{gradshteyn} we have
\beq
\fl
\left.\begin{array}{c}
D_{-i\lambda/2}\,\left(\xi\sqrt2\,\e^{\,\pi i/4}\right)\ \sim\
(2\xi^2)^{\,-i\lambda/4}\,\e^{\pi\lambda/8}\exp\left\{-\,i\,\xi^2/2\right\}\\
D_{-i\lambda/2-1}\,\left(\xi\sqrt2\,\e^{\pi i/4}\right)\ \sim\
O(\xi^{-1})
\end{array}\right\rbrace\qquad\quad(\,\xi\gg\lambda>0\,)\nn
\label{+asyt-infty}
\eeq
If instead $\xi\ll-\lambda$ we have either
$\xi\,\e^{\pi i/4}=|\,\xi\,|\,\e^{5\pi i/4}$ or else
$\xi\,\e^{\pi i/4}=|\,\xi\,|\,\e^{-3\pi i/4}\,.$
Now, for ${\rm arg}(\xi\,\e^{\pi i/4})=5\pi i/4\,,$ no reliable
asymptotic expansion is available, so that from eq.~{\bf 9.246} 3.,
p. 1094 of ref.~\cite{gradshteyn} we obtain the {\it bona fide}
leading behaviour for $\xi\ll\,-\,\lambda<0\,:$ namely,
\beq
&& D_{\,-i\lambda/2-1}\,\left(\xi\sqrt2\,\e^{\pi i/4}\right)\ =\
D_{\,-i\lambda/2-1}\,\left(|\,\xi\,|\sqrt2\,\e^{-3\pi i/4}\right)\nonumber\\
&& \sim\ \frac{\sqrt{2\pi}}{\Gamma(1+i\lambda/2)}\,(2\xi^2)^{\,i\lambda/4}\,
\exp\left\{-\,\frac{\pi\lambda}{8}+{i\xi^2\over 2}\right\}\,,
\label{+asyt+infty}\\
&& D_{\,-i\lambda/2}\,\left(|\,\xi\,|\sqrt2\,\e^{-3\pi i/4}\right)\ \sim\
(2\xi^2)^{\,-i\lambda/4}\,
\exp\left\{-\,\frac{3\pi\lambda}{8} - {i\xi^2\over 2}\right\}\nonumber\,.
\eeq
Of course, the situation becomes exactly time-reversed for the two other
linearly independent solutions:
namely, for $\xi\gg\lambda>0$ we find
\beq
&& D_{-i\lambda/2-1}\,\left(-\xi\sqrt2\,\e^{\pi i/4}\right)\ =\
D_{\,-i\lambda/2-1}\,\left(\xi\sqrt2\,\e^{-3\pi i/4}\right)\nonumber\\
&& \sim\ \frac{\sqrt{2\pi}}{\Gamma(1+i\lambda/2)}\,(2\xi^2)^{\,i\lambda/4}\,
\exp\left\{-\,\frac{\pi\lambda}{8}+{i\xi^2\over 2}\right\}\,,\nonumber\\
&& D_{-i\lambda/2}\,\left(-\xi\sqrt2\,\e^{\pi i/4}\right)\ =\
D_{\,-i\lambda/2}\,\left(\xi\sqrt2\,\e^{-3\pi i/4}\right)\nonumber\\
&& \sim\ (2\xi^2)^{\,-i\lambda/4}\,
\exp\left\{-\,\frac{3\pi\lambda}{8} - {i\xi^2\over 2}\right\}\,,
\label{-asyt+infty}
\eeq
whereas for $\xi\ll -\lambda<0$ we obtain
\beq
\fl
\left.\begin{array}{c}
D_{-i\lambda/2}\,\left(|\xi|\sqrt2\,\e^{\,\pi i/4}\right)\ \sim\
(2\xi^2)^{\,-i\lambda/4}\,\e^{\pi\lambda/8}\exp\left\{-\,{i\,\xi^2/2}\right\}\\
D_{-i\lambda/2-1}\,\left(|\xi|\sqrt2\,\e^{\pi i/4}\right)\ \sim\
O(\xi^{-1})
\end{array}\right\rbrace\qquad\quad(\,\xi\ll-\,\lambda<0\,)\nn
\label{-asyt-infty}
\eeq
For a given particle momentum $p_x=p\,,$
we shall associate the stationary asymptotic phase
\beq
px\ -\ \frac12\,{\xi^2}(t)\ =\ px\ -\ \frac12\,eEt^2 + pt - {p^2\over 2eE}
\eeq
to the {\sl positive frequency solutions} ${\xi^2}(t)$
which describe a particle, {i.e.} an electron of momentum $p\,,$
while the stationary asymptotic phase
\beq
px\ +\ \frac12\,{\xi^2}(t)\ =\ px\ +\ \frac12\,eEt^2 - pt + {p^2\over 2eE}
\eeq
will describe an antiparticle, {i.e.} a positron of momentum $-\,p\,.$

%

\section*{References}
\end{document}